\documentclass[twocolumn]{aa}

\usepackage{graphicx}
\usepackage{txfonts}
\usepackage{graphicx}
\usepackage{txfonts}
\begin{document}

\title{ A search for  spectroscopic binaries in the Galactic Globular Cluster M4\thanks{Based on
 observations  collected at  ESO  Paranal  Observatory within  the
observing program 71.D-0205 and 77.D-0182} }

\subtitle{Based on 5973 individual spectra collected at VLT.}

\author{
  V.\ Sommariva\inst{1,2}, G.\ Piotto\inst{2}, M.\ Rejkuba\inst{1}, 
  L.\ R.\ Bedin\inst{3},  D.\ C.\ Heggie\inst{4}, 
  R.\ D.\ Mathieu\inst{5}, S. \ Villanova\inst{6}
}
  \offprints{V. Sommariva}
\institute{
ESO,   Karl-Schwarzschild-Strasse  2,   D-85748 Garching bei M\"unchen,  Germany\\
\email{[vsommari;mrejkuba]@eso.org} \and
Astronomy  Department, Padova University,  vic. Osservatorio  2, 35122
Padova, Italy\\
\email{piotto@unipd.it} 
\and
Space Telescope  Science Institute, 3700 San  Martin Drive, Baltimore,
MD, USA\\
\email{bedin@stsci.edu} 
\and
School of Mathematics and Maxwell Institute for Mathematical Sciences,
University of Edinburgh, Edinburgh, EH9 3JZ, UK\\
\email{d.c.heggie@ed.ac.uk} 
\and
Department of Astronomy, University of Wisconsin  Madison, WI, 53706, USA\\
\email{mathieu@astro.wisc.edu}
\and
Grupo   de  Astronomia,   Departamento  de   Fisica,   Universidad  de
Concepcion, Casilla 160-C, Concepcion, Chile.\\
\email{svillanova@astro-udec.cl} 
}

\date{Received 2008 July, Accepted  2008 October 10.}
\titlerunning{Spectroscopic binaries in M4}
\authorrunning{V. Sommariva et al.}

% \abstract{}{}{}{}{} 
% 5 {} token are mandatory

\abstract
% context heading (optional)
% {} leave it empty if necessary  
{}
% aims heading (mandatory)
{We present a large multi-epoch high resolution spectroscopic investigation
for the search of binary candidates in the Galactic Globular Cluster (GGC) M4. 
The aim of our work is the identification of the binary candidates, and 
the determination of the binary fraction and of the binary
radial distribution.}
% methods heading (mandatory)
{We have obtained 5973 individual spectra of 2469 
stars with the multifiber facility FLAMES+GIRAFFE at VLT.
 Selected stars go
from the red giant branch tip to one magnitude below the TO, and cover the
entire cluster. Each star has
been observed at least twice, in a temporal interval of about
three years. For 484 stars we have three epochs. Radial velocities have been obtained with the
classical cross-correlation technique, using a solar type spectrum as template.}
% results heading (mandatory)
{The average
radial velocity of the observed cluster members is $70.29 \pm 0.07 (\pm 0.3)(\pm0.1)$~km~s$^{-1}$.

The search for variations in radial velocities among the stars with multi-epoch
observations yielded 57 binary star candidates.\\
Our radial velocity measurement accuracy allowed us to identify  at a $3\sigma$ level binaries 
with radial velocity variations larger than $\sim 0.3$~km~s$^{-1}$ for
the target stars with $V\leq15$, and larger than $\sim 0.5$~km~s$^{-1}$ for the targets
with $V\ge15$.
We identified 4 binary star candidates out of 97 observed targets inside the core
radius, and 53 candidates out of 2372 observed stars outside the core radius. 
Accounting for the incompleteness affecting our survey, 
the lower limit for the total binary fraction is  $f=3.0\pm0.3\%$.
The lower limit for the binary fraction in the cluster core is 
$f=5.1\pm2.3\%$, while outside the core it decreases to 
$f=3.0\pm0.4\%$. 
Similarly, we found $f=4.5\pm0.4\%$ and $f=1.8\pm0.6\%$ for the binary fraction inside 
and outside the half mass radius.}
% conclusions heading (optional), leave it empty if necessary 
{}

\keywords{spectroscopy -- 
          globular clusters: individual: M4 (NGC~6121) -- 
          binaries
          }

\maketitle
%
%________________________________________________________________

%%%%%%%%%%%%%%%%%%%%%%
%
\section{Introduction}
%
%%%%%%%%%%%%%%%%%%%%%%

The frequency of binary systems is one of the key 
parameters for dynamical models of star clusters 
(see review by Hut et al. 1992, and references therein). 
Through their formation and 
destruction, binaries play a fundamental role in the dynamical evolution 
of a globular cluster, especially during the core collapse phase. 
In fact they can be seen as an efficient
heating source that can halt the core collapse 
(Hut, McMillan, \& Romani, 1992).
Additionally, encounters between binary systems with other single or 
multiple stars, can on the one hand disrupt the wider binaries, and on the other 
lead to the formation 
of some exotic objects like blue stragglers, millisecond pulsars or x-ray binaries
(e.g.\ Bailyn 1995).
There are three basic search techniques for binary systems in a star cluster: 
(i) photometric study of the so-called secondary sequence in the 
color-magnitude diagram created by the superposition of two main sequence stars 
(e.g.\ Bellazzini et al. 2002, Sollima et al. 2007, Milone et al. 2008); 
(ii) the study of the light curves of eclipsing binaries (e.g.\ Yan \& Mateo 1994, 
Albrow et al.\ 2001, Kaluzny et al.\ 2008); and (iii) the
search for  spectroscopic radial velocity  variations of the stars at different 
epochs (e.g.\ C\^ot\'e \& Fischer 1996, Pryor et al.\ 1988, C\^ot\'e et al.\ 1996). 
All three methods have some advantages and some limitations.
For instance, the photometric study has the advantage of needing only one 
observational campaign, but it cannot constrain the orbital parameters. 
On the other hand, the search for eclipsing binaries is sensitive only to systems 
with periods less than 10 days (Mateo 1993), and moreover it has some limitations 
given by the inclination of 
the systems, and it requires  multi-epoch observations. The latter limitation is
common to the spectroscopic search for velocity variations. However, this method has 
important advantages over the last two, because it allows to determine 
orbital period and the eccentricity in an unbiased way. 
Observational constraints of the period distribution for binary systems 
are interesting, not only because dynamical processes preferentially 
eliminate the wider systems (Heggie 1975, Hut 1983),
but also because when the period is known it is possible to discriminate
between tidal capture and primordial binaries and give some constraints on the
system's mass.\\
Until now there have been few searches for binary stars in globular clusters 
through spectroscopic observations of radial velocity variations. 
In particular, most of the previous work has been limited either to bright
red giant branch stars (e.g.\ M22 observations by C\^ot\'e et al. 1996), or, 
due to the relatively low multiplex capabilities of instruments, to a relatively small 
number of stars per cluster (e.g.\ Yan \& Cohen 1996; C\^ot\'e \& Fischer 1996). 
A list of all the published spectroscopic binary star investigations is 
shown in Table 1: the first column gives the name of the globular cluster,
the second column the number of stars considered, the third one the number
The most complete spectroscopic binary-search in 
M4 is by C\^ot\'e  \& Fischer  (1996), who analyzed  33 turn-off dwarf stars. Their 
observations were tuned to search for ultra-hard binaries on the main sequence 
of the cluster. From the observed variability of two stars, they conclude that 
the best-fit binary fraction in this cluster is $15\pm15\%$.
We didn't find any stars in common with their.
In this work we analyze 
several thousand stars in order to estimate of 
the binary fraction in M4 on a firmer ground .

The reasons to select M4 as our target are the following.
First of all, it shows no evidence for any central brightness cusp 
(Trager et al.\ 1995) indicative of core collapse, 
despite the fact that the cluster age exceeds its 
central relaxation time  (Harris 1996)
by a factor of approximately 400. Is the lack of evidence of core collapse due to the
presence of a large fraction of binaries? 
And let us compare M4 with NGC6397. NGC6397 has a mass, relaxation time and
Galactic orbit similar to those of M4. 
Therefore it should be at a similar stage of its evolution, 
and yet it has  a completely different structure with a collapsed core. 
Is this due to the lack of primordial binaries?
The aim of our investigation is to test this theoretical picture 
for the first time against observations.
This is a fundamental test of our current understanding of the problem, and
vital input for future theoretical research. In fact, the advent of dedicated machines
like the GRAPEs (Hut \& Makino 1999), already 
allows realistic N-body modelling of (open) star clusters with up 
to $5\times10^{4}$ particles  (Portegies Zwart et al.\ 2004).
In a few  years from now, with GRAPE-DR, realistic
N-body modeling of GC with up to $5\times10^{5}$ particles (like the 
estimated original number of stars in M4 and NCG6397, which makes these two clusters 
the ideal test beds) will be feasible.
At that point, observational inputs, and the comparison of the model with the 
observed parameters will become of fundamental importance for our understanding of
the evolution of the GC, similarly to what
happens when we compare stellar evolution models with observed 
color-magnitude diagrams and luminosity functions. 

In order to make realistic models, a number of observational data are needed. 
Most of them are already available for M4, including mass functions 
at different 
radial distances, the proper motion  and the radial velocity information.
This work provides significant improvement to the latter.
Only the fundamental information on its binary population is missing. 
Additional reasons which make M4 our first priority target are that it has an extended 
(relatively uncrowded) core, making it feasible the search for binaries with a 
multifiber facility as FLAMES@VLT where the portion  of binaries should be highest. 
Moreover, M4 is the cluster closest to the Sun,
which enables us to study the binaries below the turn-off, and these
are pristine tracers, as binaries containing giants might be destroyed by 
internal mass transfer (a prediction that we want to check by observing a few hundred giants).
The ultimate goal of the present study is to establish
the frequency of binary systems in M4, and to determine their radial distribution 
within the cluster. These two quantities are intimately linked to the 
internal dynamics of the cluster. 
Comparing the current binary frequency in the cluster with the results of dynamical
modeling, it is then possible to infer important constraints on the primordial
binary population.

Heggie \& Giersz (2008) recently constructed a dynamical evolutionary model for M4.
It is based on a Monte Carlo simulation including an appropriate stellar initial mass 
function, a primordial binary population, galactic tidal effects, 
synthetic stellar and binary evolution, relaxation, and three- and four-body 
interactions.
 The most surprising discovery from this model was that M4 
was found to be a post-collapse cluster sustained by binary burning. 
The model provides also detailed predictions for the period and spatial 
distribution of binary systems.

In this paper we will present the results of the spectroscopic search 
for binary systems in M4. In a forthcoming
next publication  we will make detailed comparisons 
of our results with the model of Heggie \& Giersz (2008). 

The organization of this work is as follows: the observations and sample 
selection are presented in Section 2, 
the data reduction is described in Section 3,
the radial velocity analysis is carried out in Section 4, 
while the results are presented in Section 5.
The spatial distribution and the lower limits on the 
frequency of binaries are discussed in the last section.

\begin{table*}
\caption{Spectroscopic binary star investigations carried out till now.}
\begin{center}

\begin{tabular}{lcccccccccccc} \hline
\hline
Globular Cluster ID             & number target & number binary     & reference  	\\ 
\hline
\hline
M3			      &  111  &  0  &   Gunn \& Griffin (1979)\\
M3			      &  111  &  1  &   Pryor et al.	(1988)\\
47 Tuc,M2,M3,M12,M13,M71 			      &  393  &  6  &   Pryor et al.    (1989)\\
NGC 3201		      &  276  &  2  &	C\^ot\'e et al.     (1994) \\
NGC 5053		      &   66  &  6  &   Yan \& Choen    (1996)\\
M71			      &	 121  &	12  &	Barden et al.   (1996)\\
M22			      &  109  &  1  &	C\^ot\'e et al.     (1996)\\
M4			      &	  33  &	 2  &	C\^ot\'e et al.     (1996)\\
Pal 5			      &	  18  &	 1  &	Odenkirchen et. al. (2002)\\
NGC 6752		      &	  51  &	 0  &	Moni Bidin et al. (2006)\\
\hline
\hline
\end{tabular}
\end{center}
\end{table*}

%__________________________________________________________________

%%%%%%%%%%%%%%%%%%%%%%
%
\section{Description of the observations}
%
%%%%%%%%%%%%%%%%%%%%%%

In July 2003, we collected a single spectrum for 2684 stars in
the GGC M4.  The main purpose of this original project (ESO-program
{\sf 71.D-0205}) was the study of the internal velocity dispersion
field of M4. In the following, we will refer to
these observations as the first epoch (epoch~I).
For 2469  stars observed in the first epoch  we  obtained  
a second  spectrum  between 
July and September 2006 (epoch~II)  under ESO-program {\sf 77.D-0182} with
the aim of measuring radial velocity variations for those stars in
order to identify binary candidates.

In addition  to these, for  a subsample of  364 stars, mainly
in the core of the cluster,  we  collected a
third epoch (epoch~III), separated by  $\sim5$ weeks from epoch~II, 
with the intent of detecting short-period binaries.

All  data   were  obtained  with  the   high  resolution  spectrograph
FLAMES+GIRAFFE at VLT (Pasquini  et al.\ 2000).  FLAMES is  the Fibre Large
Array Multi Element Spectrograph mounted  at the Nasmyth A platform of
the  8.2m Kueyen  (UT2) telescope,  which is  part of  the  Very Large
Telescope (VLT) of the European Southern Observatory (ESO) situated on
Cerro  Paranal.  With FLAMES  in MEDUSA
 mode it is possible to observe up to 132 
targets at the same time over a field of view of 25 arcmin diameter.

The fibers are accurately placed with the aid of magnetic buttons on a
metallic plate (two plates are available  at any time: one
in place, and the other preparing the buttons for the next pointing).
During epoch~I  twenty five plates  were collected, during  epoch~II in
addition to the same number  of plates, three plates with stars placed
mainly in the  core were taken. These three  plates have been repeated
during epoch~III.

The GIRAFFE spectrograph  operates at resolutions  R$=$6,000-48,000 across
the  entire visible  range,  360$-$940  nm. It  is  equipped with  two
gratings,  high  (HR)  and  low   (LR)  resolution  and  with  a  single
2K$\times$4K EEV CCD (15 $\mu$m pixels).

FLAMES is the ideal instrument for  this kind of study thanks to the
large field of  view, the high multiplex capability,  the precision in
radial velocities and the faint magnitudes that it can reach.
For our project the ability to register the radial velocities of stars
into a common reference system  is a crucial issue.  Any unexpected
systematic  error  between epochs  would  introduce  serious limits to
the accuracy of the radial velocity variations that we could see. 
For this  reason during  our observations we  implemented simultaneous
calibrations using the Th-Ar lamp.
In total  5973 spectra were  collected in three different  epochs from
the red giant tip to below the Turn$-$Off (TO).
Our sample  covers most of the cluster  extension, from the inner
core out to the cluster outskirts.   For 2469 stars we have spectra in
two epochs; for  a subsample of 484 stars, we have
three epochs. For 34 stars we obtained also a fourth epoch.
This is due to  the fact that together with the 364  stars in the core
for  which we  obtained the  third  epoch (three repeated plates),  we
observed an additional  120 stars randomly selected among the targets of
epoch~II.
These  repeated  stars  will  provide important  cross-checks  of  the
registration of different plates to a common radial velocity system.

Epochs~I and  II have the  purpose of detecting 
mainly 
soft-binary candidates
with periods up to few years from the cluster center to its outskirts.
Epoch~III  was taken  a few weeks  after  epoch~II with  the purpose  of
identifying  hard binaries  with periods  up to  a few  weeks  inside the
cluster core, where  binaries are expected to be  more numerous because of
mass-segregation.

All stars, at all epochs, were selected from our astrometric and
photometric catalog (Anderson et al.\ 2006) with the following
criterion: each star at any given magnitude $V$ had no neighbors
brighter than a magnitude $V+2.5$ within an angular distance of 1.2
arcsec (for comparison, the radius of the fibre 
correspond to only 0.6 arcsec).
The catalogue is based on Wide Field Imager (WFI) data from the ESO/MPIA
 2.2m telescope, and has astrometric internal precision of $\sim7$mas.
 More recently the photometry has been revised by Momany 
 (private communication). Therefore we decided to use
that photometry in the following.
The  target  stars  are  shown  as thick (red) dots in the
color-magnitude diagram presented in Fig.~1.

All stars  were observed with HR9 setup  (514.3-535.6nm) centered at
525.8nm.  This  instrumental configuration offers high resolution
($R=25,800$)  and the  best radial  velocity accuracy  (Royer  et al.\
2002). This setup  covers 200 \AA, and a CCD  pixel corresponds to 0.05
\AA.   It provides  the best  compromise in  terms  of signal-to-noise
ratio (SNR) achievable for red giant stars and the number
of spectral lines  necessary  for accurate Doppler measurements.
The stars were grouped in three different subsample centered at $V=$16.5,
17, and 17.5, and their spectra were collected with an exposure time
of 1200s, 1500s, and 1800s, respectively, in order to guarantee that each spectrum
has a minimum SNR=10.  On the basis of our experience with previous M4
observations and the GIRAFFE pipeline, this is the minimum signal to noise
ratio that we need to reach the required radial velocity precision
of a few hundreds m/s.
Figure 2 shows the positions of the target stars with respect to the
cluster center.
In our observational strategy we covered all the isolated bright stars
in the FLAMES field of view.
%%%%%%%%%%%%%%%%%%%%%%%%%%%%%%%%%%%%%%%%%%%%%%%%%%%%%%%%%%%%%%%%%%
%%%%%%%%%%%%%%%%%%%%%%%%%%%%%%%%%%%%%%%%%%%%%%%%%%%%%%%%%%%%%%%%%%
%%%%%%%%%%%%%%%%%%%%%%%%%%%%%%%%%%%%%%%%%%%%%%%%%%%%%%%%%%%%%%%%%%
%%%%%%%%%%%%%%%%%%%%%%%%%%%%%%%%%%%%%%%%%%%%%%%%%%%%%%%%%%%%%%%%%%

\begin{figure}[!ht]
\resizebox{\hsize}{!}{ \includegraphics{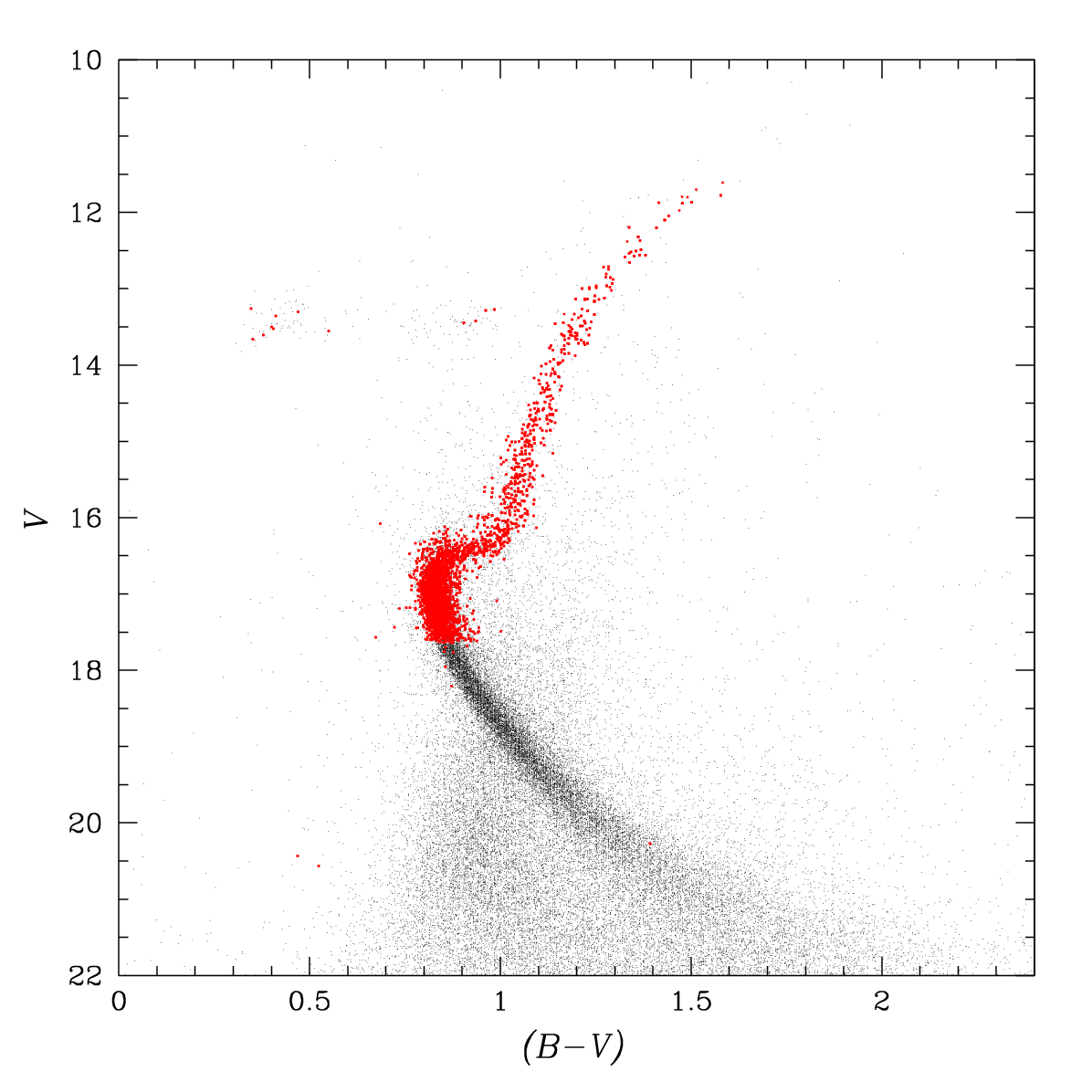}}
\caption[]{
Small dots show  the color-magnitude diagram for all  the stars in our
astro-photometric catalog.  The stars  highlighted with 
thick dots 
(in red  in  the  electronic version)  are  the  stars  for which  we  have
collected spectra. 
}
\label{CMD}
\end{figure}

Table 3  contains the list of  all  the plates  used for  the
present work. The first column lists the plate ID (which, according to
ESO file naming convention, contains also the time and date of the 
exposure start), the second column
contains  the exposure times, and the  third one gives the value  of the
air mass during the observation.  The last column reports the seeing 
as measured by DIMM, the ESO seeing monitor in Paranal.

Figure 3 displays the value of the signal to noise ratio for all our
spectra as a function of the visual magnitude.  To estimate the SNR we
selected three regions of the spectrum dominated by the continuum, and
took the average of the three vales of SNR determined with the
IRAF\footnote{IRAF (Image Reduction and Analysis Facility) is
distributed by the National Optical Astronomy Observatories, which
are operated by the Association of Universities for Research in
Astronomy, Inc., under cooperative agreement with the National Science
Foundation.}  {\sf splot} task.  Inspection of the spectra confirms
these values to be consistent with the Poisson noise associated with
object$+$sky signal.  In the figure the asterisks indicate the stars
observed with 1200s exposures, the squares the stars observed with
1500s exposures, and the three point stars the stars observed with
1800s exposure.

\begin{table}[!ht]
\caption[]{The fundamental parameters of M4. }
$$ 
\begin{array}{p{0.5\linewidth}ll}
\hline
\hline
\noalign{\smallskip}
Parameter      &  Value & Ref. \\
\noalign{\smallskip}
\hline
\noalign{\smallskip}
($\alpha$,$\delta$)$_{\rm J2000.0}$ & (16^{h}23^{m}35.5^{s}, -26^{\circ}31' 31'' ) & 1\\
($\ell$,$b$)$_{\rm J2000.0}$        & (350^{\circ}.97, 15^{\circ}.97)              & 1\\
distance                            & 1.7 ~ {\rm kpc}                              & 2\\
core radius r$_{\rm c}$             & 0.53 {\rm pc}                                & 3\\
half mass radius r$_{\rm h}$        & 2.3  {\rm pc}                                & 3\\
tidal radius			    & 21 {\rm pc} 				   & 3\\	
$\rm [Fe/H]$                        & -1.07                                        & 4\\
Present-day Total Mass              & 63,000 {\rm M_\odot}                         & 5\\
Age                                 & 12 {\rm Gyr}                                 & 6\\
Radial velocity 		    & 70.29 ~ {\rm km s^{-1}}			   & 7\\
\noalign{\smallskip}
\hline
\end{array}
$$ 
\begin{list}{}{}
\item[$^{\mathrm{1}}$] Djorgovski \& Meylan (1993)
\item[$^{\mathrm{2}}$] Peterson et al.\ (1995) 
\item[$^{\mathrm{3}}$] Trager et al. \ (1993)
\item[$^{\mathrm{4}}$] Marino et al. \ (2008)
\item[$^{\mathrm{5}}$] Richer et al.\ (2004)
\item[$^{\mathrm{6}}$] Hansen  et al.\ (2004) 
\item[$^{\mathrm{7}}$] This work 
\end{list}
\label{M4par}
\end{table}
%__________________________________________________ 
%

%%%
%   FIGURE 
%__________________________________________________ 
%
%__________________________________________________ 
%

%%%
%   FIGURE 
%__________________________________________________ 
%
\begin{figure}[!ht]
\resizebox{\hsize}{!}{ \includegraphics{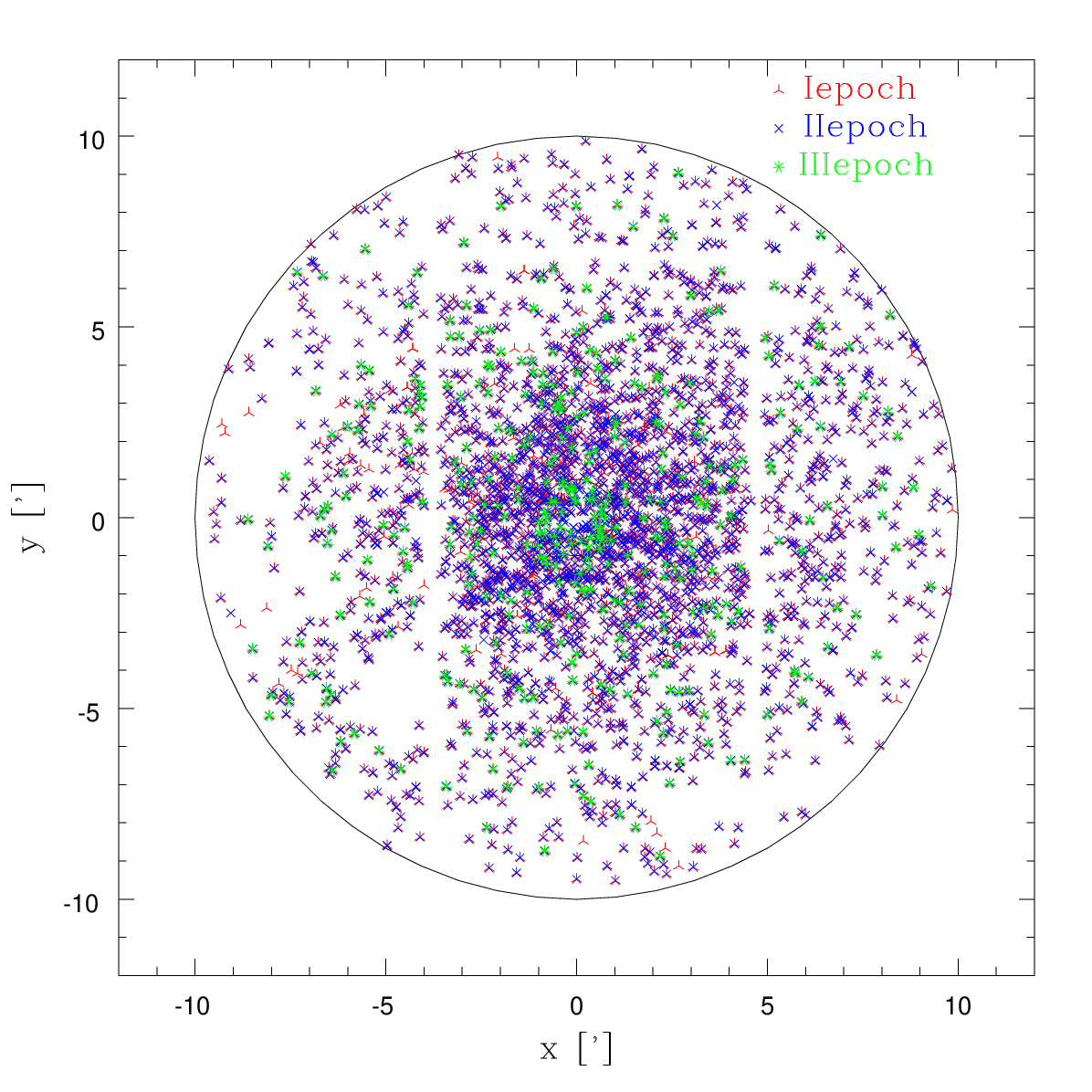}}
\caption[]{
Spatial distribution  of the target stars with respect to the center reported
in Tab.~1.  The
coordinates are given in arcminutes.
Observations from different epochs are shown with different symbols (and
colors).
The core radius is 0.83, corresponding to 0.53pc assuming 
a distance of 1.7 kpc for M4.
The white stripes are due 
to gaps between the CCDs of the WFI at 2.2m camera where our astrometry 
comes from.
}
\label{Fib}
\end{figure}
%__________________________________________________ 

%   FIGURE 
%__________________________________________________ 
%
\begin{figure}[!ht]
\resizebox{\hsize}{!}{ \includegraphics{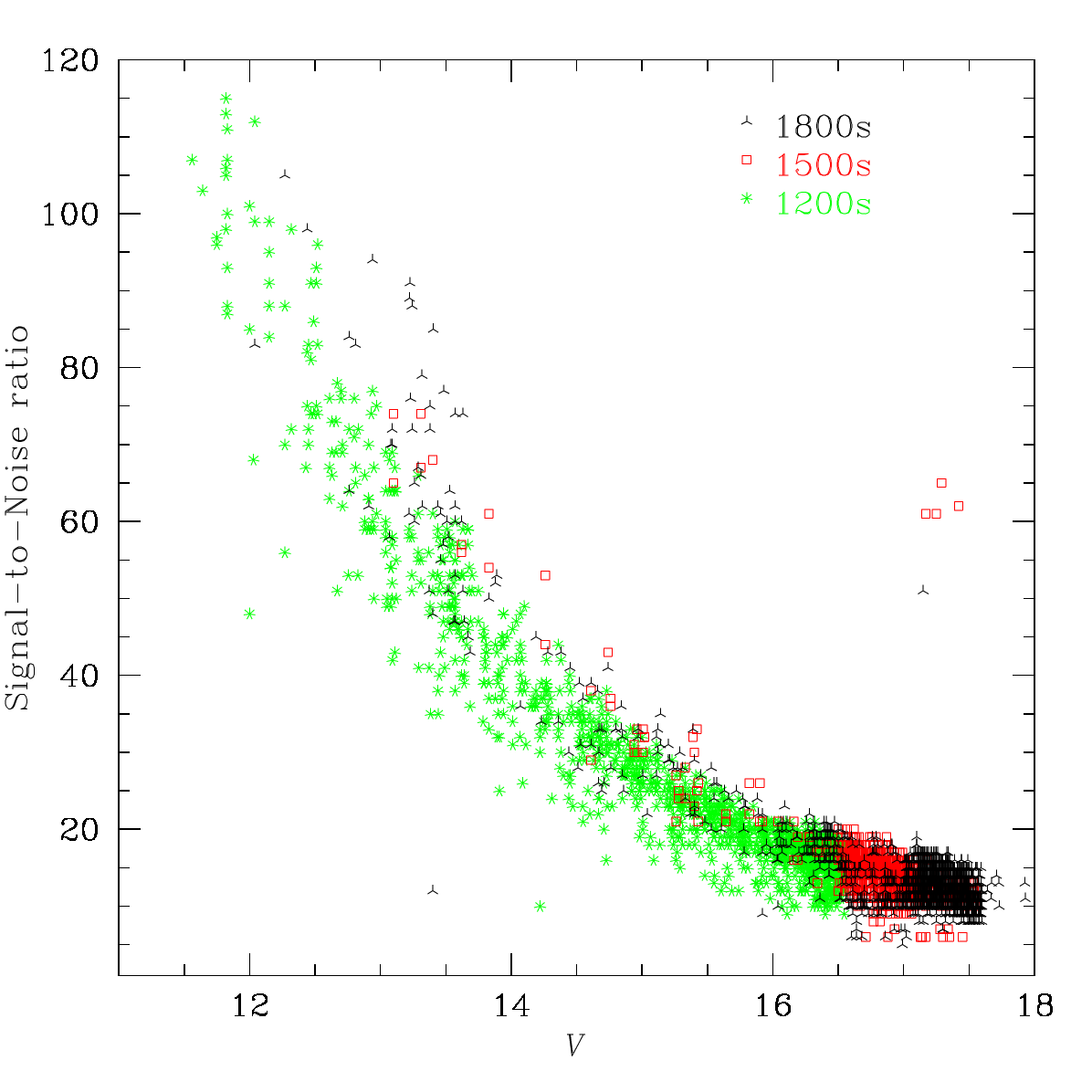}}
\caption[]{
Estimated signal-to-noise ratio for each of our spectra (see text).
Different symbols indicate stars observed 
with different exposure times.
}
\label{SNR}
\end{figure}
%__________________________________________________ 
%

\begin{table}
%\begin{minipage}{\textwidth}
\caption{
Summary of the observations. 
Repeated plates between epoch~II and epoch~III 
are marked with an $*$. 
}

\begin{tabular}{llll} 
\hline
\hline
Plate ID                & Exp.     & airmass  & Seeing  	\\ 
$[$archive name$]$      & [s]      & sec($z$) & [$''$]        	\\
\hline
\multicolumn{4}{c}{epoch I}\\
\hline
GIRAF.2003-06-01T01:55:28 &  1200 &   1.213 &  	0.8 \\
GIRAF.2003-06-29T02:49:33 &  1200 &   1.002 &  	0.6 \\
GIRAF.2003-07-02T05:11:52 &  1200 &   1.256 &  	1.3 \\	
GIRAF.2003-07-09T04:27:48 &  1200 &   1.200 &  	0.7 \\
GIRAF.2003-07-09T04:59:48 &  1200 &   1.318 &  	0.6 \\
GIRAF.2003-07-09T05:41:10 &  1200 &   1.548 &  	0.7 \\
GIRAF.2003-07-09T06:22:54 &  1200 &   1.935 &  	0.6 \\
GIRAF.2003-07-20T04:56:23 &  1200 &   1.538 &   0.6 \\					     	     
GIRAF.2003-07-08T03:32:29 &  1500 &   1.066 &  	0.7 \\
GIRAF.2003-07-08T04:09:32 &  1500 &   1.139 &  	0.7 \\
GIRAF.2003-07-08T04:51:48 &  1500 &   1.269 &  	0.6 \\
GIRAF.2003-07-08T05:31:32 &  1500 &   1.461 &  	0.7 \\
GIRAF.2003-07-08T06:13:11 &  1500 &   1.785 &  	0.7 \\
GIRAF.2003-07-21T04:07:03 &  1500 &   1.295 &   0.5 \\
GIRAF.2003-07-22T03:49:21 &  1500 &   1.242 &  	0.5 \\
GIRAF.2003-07-22T04:24:29 &  1500 &   1.394 &  	0.5 \\					     	     
GIRAF.2003-07-07T04:25:37 &  1800 &   1.170 &  	1.3 \\
GIRAF.2003-07-20T01:42:22 &  1800 &   1.006 &  	0.9 \\
GIRAF.2003-07-20T03:33:25 &  1800 &   1.167 &   0.9 \\
GIRAF.2003-07-20T04:16:24 &  1800 &   1.318 &  	0.7 \\
GIRAF.2003-07-21T01:54:33 &  1800 &   1.015 &  	0.7 \\
GIRAF.2003-07-21T02:40:59 &  1800 &   1.066 &   0.8 \\
GIRAF.2003-07-21T03:25:57 &  1800 &   1.158 &  	0.4 \\
GIRAF.2003-07-21T04:50:01 &  1800 &   1.522 &  	0.6 \\
GIRAF.2003-07-22T02:02:22 &  1800 &   1.024 &  	0.5 \\
\hline
\multicolumn{4}{c}{epoch II}\\
\hline
$*$ 
GIRAF.2006-07-31T01:34:42 &  1200 &   1.032 &  	0.8 \\
GIRAF.2006-09-04T23:24:34 &  1200 &   1.045 &  	0.8 \\
GIRAF.2006-09-04T23:59:28 &  1200 &   1.101 &  	0.9 \\
GIRAF.2006-09-05T01:24:01 &  1200 &   1.380 &  	1.0 \\
GIRAF.2006-09-05T00:50:39 &  1200 &   1.238 &  	0.9 \\
GIRAF.2006-09-06T00:53:35 &  1200 &   1.264 &   0.8 \\
GIRAF.2006-09-06T02:10:30 &  1200 &   1.728 &  	0.8 \\
$*$ 
GIRAF.2006-07-31T02:09:45 &  1500 &   1.080 &   0.8 \\
GIRAF.2006-08-01T23:28:08 &  1500 &   1.025 &  	0.9 \\
GIRAF.2006-08-02T01:12:13 &  1500 &   1.019 &  	0.8 \\	
GIRAF.2006-08-02T01:52:06 &  1500 &   1.064 &  	0.7 \\
GIRAF.2006-08-02T02:27:43 &  1500 &   1.132 &  	0.7 \\
GIRAF.2006-08-02T03:03:13 &  1500 &   1.234 &  	0.8 \\
GIRAF.2006-09-04T01:04:04 &  1500 &   1.274 &   0.8 \\
GIRAF.2006-09-04T00:20:55 &  1500 &   1.139 &   0.8 \\
GIRAF.2006-09-05T02:01:48 &  1500 &   1.620 &  	0.9 \\
$*$ 
GIRAF.2006-07-31T02:48:50 &  1800 &   1.166 &   0.8 \\
GIRAF.2006-07-31T03:38:40 &  1800 &   1.346 &   0.7 \\
GIRAF.2006-08-10T00:25:57 &  1800 &   1.009 &   1.4 \\ 
GIRAF.2006-09-03T23:40:14 &  1800 &   1.061 &   1.6 \\
GIRAF.2006-09-06T00:11:51 &  1800 &   1.137 &  	0.6 \\
GIRAF.2006-09-06T01:28:52 &  1800 &   1.427 &  	1.0 \\
GIRAF.2006-09-07T01:08:31 &  1800 &   1.343 &  	1.0 \\
GIRAF.2006-09-14T23:55:17 &  1800 &   1.187 &  	1.1 \\
GIRAF.2006-09-15T23:54:20 &  1800 &   1.196 &   1.2 \\
GIRAF.2006-09-16T23:56:54 &  1800 &   1.216 &  	0.9 \\
GIRAF.2006-09-17T00:37:45 &  1800 &   1.384 &  	1.0 \\
GIRAF.2006-09-18T00:02:53 &  1800 &   1.251 &  	0.8 \\
\hline
\multicolumn{4}{c}{epoch III}\\
\hline
$*$ 
GIRAF.2006-09-05T23:39:26 & 1200  &   1.394 &   1.3 \\
$*$ 
GIRAF.2006-09-05T02:37:05 & 1500  &   1.072 &   1.1 \\
$*$ 
GIRAF.2006-09-06T23:45:30 & 1800  &   1.190 &   1.0 \\
\hline
\end{tabular}
\label{plates}
\end{table}

%%%%%%%%%%%%%%%%%%%%%%%%%%%%%%%%%%%%%%%%%%%%%%%%%%%%%%%%%%%%%%%%%%%%%%
%%%%%%%%%%%%%%%%%%%%%%%%%%%%%%%%%%%%%%%%%%%%%%%%%%%%%%%%%%%%%%%%%%%%%%
%%%%%%%%%%%%%%%%%%%%%%%%%%%%%%%%%%%%%%%%%%%%%%%%%%%%%%%%%%%%%%%%%%%%%%
\section{Data reduction}
\label{RED}
%%%%%%%%%%%%%%%%%%%%%%%%%%%%%%%%%%%%%%%%%%%%%%%%%%%%%%%%%%%%%%%%%%%%%%
%%%%%%%%%%%%%%%%%%%%%%%%%%%%%%%%%%%%%%%%%%%%%%%%%%%%%%%%%%%%%%%%%%%%%%
%%%%%%%%%%%%%%%%%%%%%%%%%%%%%%%%%%%%%%%%%%%%%%%%%%%%%%%%%%%%%%%%%%%%%%

The  data reduction  was  performed using  the  pipeline developed  at
Geneva  and Paris  Observatories (Blecha  et al.\  2000):  the GIRAFFE
BaseLine Data Reduction Software (girBLDRS), version 1.13.  With it we
removed the instrumental signature from the observed data, subtracting
the  bias  and dividing  by  the  normalized flat--field.   Flat-field
observations were also  used to trace the position  of all the fibres,
and to  derive the  parameters for the  optimal extraction  of science
exposures.  Finally,  the wavelength calibration  was determined using
the day-time Th-Ar lamp exposures.

As we emphasized above, in order to to achieve our scientific goals it
is fundamental to register all the radial velocity measurements into a
common reference system free --as much as possible-- of any systematic
error down to $\sim$1 {\rm km s$^{-1}$} .

Drifts  inside the  instrument could  introduce  {\em fiber-to-fiber},
{\em  plate-to-plate}, and {\em  epoch-to-epoch} systematic  errors in
the radial velocity measurements. 
These drifts  are due to  different conditions of the  instrument (for
example  a   variation  of  the  temperature   between  the  night-time
calibrations and during the night--time  science exposures, or
small mechanical shifts caused by  earthquakes that occur rather 
frequently in northern Chile).
Such differences would result not  only in a different global shift of
the velocities derived in the  two cases, but also in a fiber-to-fiber
difference as result of a different rotation of the slit geometry.

In Sect. 4  we will deal with the  plate-to-plate and epoch-to
epoch systematic errors that can be treated as global shifts.
Most of the fiber-to-fiber correction  is achieved  by adjusting the slit geometry
using the day-time Th-Ar exposure closest in time to science observations.
This is done 
with  the  {\sf  wcalslit}  pipeline  task. The remaining fibre-to-fibre
drifts are corrected using the five simultaneous Th-Ar fibres that are
present in all our scientific frames. We  first  describe  {\sf
wcalslit}  and its application,  because it  has to  be done  prior to
final fibre extraction and wavelength calibration.

%
%----------------WCALSLIT----------------------------------------------
%
The pipeline  uses the physical  description of the slit  geometry and
therefore  iteratively,  using  the  first  guess  of  the  wavelength
solution, matches  the measured to expected positions  of the emission
Th-Ar lamp lines.  However, if there are any drifts in the instrument,
the  slit geometry  needs  to be  adjusted.  The procedure  on how  to
correct  the  slit  geometry  and  therefore  improve  the  wavelength
calibration using only the day-time Th-Ar calibration frames is given
at the  url {\sf  http://girbldrs.sourceforge.net/} by the  authors of
the pipeline.
%
%%%
%%%%
The {\sf wcalslit} task determines the positions of all the 135 fibres
in the Th-Ar day-time lamp  frame and compares them with the expected
positions in the  {\sf girSlitGeoMedusa1(2)H9.tfits} file which  is one of
the necessary  calibration files of  the pipeline. If  differences
are found, {\sf wcalslit} updates the  table.  This task has to be run
separately  for Medusa1  and Medusa2,  i.e.\ the  two plates  that are
available for FLAMES-Medusa observations.

After adjusting  the slit geometry  calibration files in this  way, we
proceeded with  the data  reduction, by re-extracting  the calibration
frames:  flat-field  and  wavelength  Th-Ar calibration,  as  well  as
science exposures.  The newly  extracted flat-fields were divided, and
newly determined wavelength calibration applied.

FLAMES   offers   the   possibility  to   observe   {\em
simultaneously} with  the on-sky  exposure also the  Th-Ar calibration
lamp  source  through  five   MEDUSA  fibres,  which  are  distributed
uniformly over  the detector:  fibres 1, 32,  63, 94, and  125.  These
five simultaneous fibres are essential  to achieve the optimal
wavelength calibration.

Therefore, in order  to  verify and  eliminate  any  residual wavelength  drifts
across the  130 fibres, we acquired  simultaneous calibration spectra
in each of  our observations.  The unfortunate drawback  in having the
simultaneous Th-Ar lamps on  during the 30min long exposures 
is some  contamination of  the stellar spectra  in the  fibres located
directly  next to  the simultaneous  fibres  on the  detector. In  the
subsequent analysis we discarded those spectra, because their velocity
measurements  had larger  errors  or  in few a  cases  of faint  stellar
spectra could not be reliably determined.

The current version of  girBDLRS corrects the variations of wavelength
solution  across  the CCD  in  an  automated  way.  After  the  fibre
extraction, and  wavelength calibration  of all the  fibres (including
simultaneous calibration fibres)  according to the solution obtained using
the reference Th-Ar day-time calibration frame, it computes the shift
between the expected and measured  position of Th-Ar lines for each of
the  five simultaneous  fibres.   For each  of  the five  simultaneous
calibration fibres, the  average shift is computed and  then a linear
regression is fitted to fibre position vs.\  average shift. Wavelength
calibrations  for  all the  fibres  are  then  corrected by applying  the
computed linear correction. In the second iteration the shifts between
the  expected and  measured average  positions of  Th-Ar lines  of the
simultaneous calibration fibres are typically smaller than 0.001 nm 
(i.e. $\sim$0.6  {\rm km s$^{-1}$}).

%%%

In total we  had 56 frames, each with  80$-$130 spectra, including the
sky and the simultaneous calibration fibers.  At the end we obtained a
total of 5973 stellar spectra.
In Figure 4 we show  examples of the spectra that we obtained after
the reduction for bright, intermediate and faint stars ({\it V} = 12,
14, and 16, respectively).

\begin{figure}[!ht]
\resizebox{\hsize}{!}{ \includegraphics{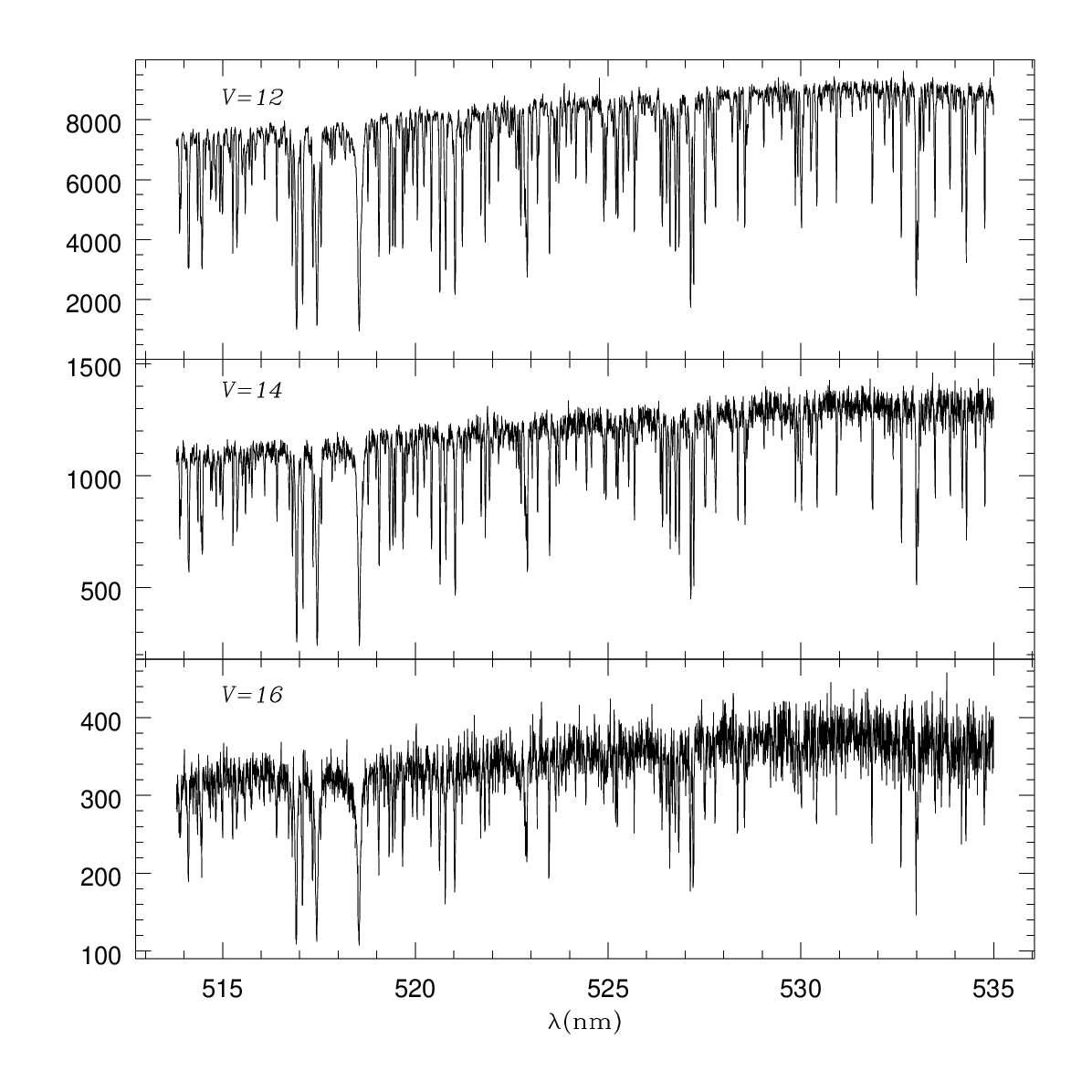}}
\caption[]{
Extracted spectra for three stars with different magnitudes:
in the upper panel for a bright star with {\it V} magnitude 12,
in the middle panel for an intermediate star with   {\it V} magnitude 14,
and in the bottom panel for a faint star  with {\it V} magnitude 16.
}
\end{figure}

\begin{figure}[!ht]
\resizebox{\hsize}{!}{
\includegraphics{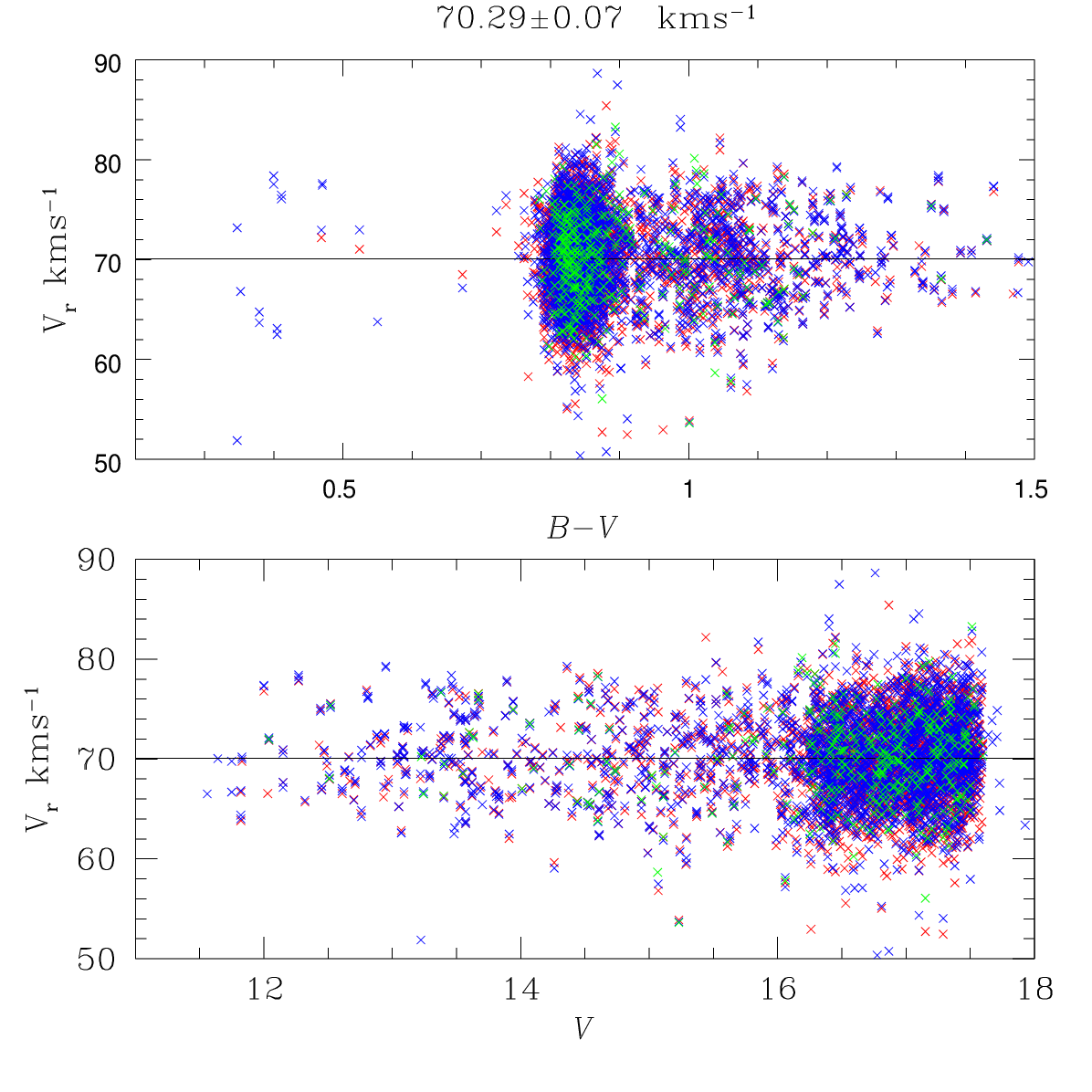}
}
\caption[]{ Radial velocities as a function of color
 (upper panel) and magnitude (bottom panel) obtained for our individual spectra. 
            Different colors refer to different epochs.
	The horizontal line indicates the mean radial velocity (value at the top)
	of all the stars.  
             }
\label{VrBV_V}
\end{figure}
%%%%%%%%%%%%%%%%%%%%%%%%%%%%%%%%%%%%%%%%%%%%%%%%%%%%%%%%%%%%%%%%%%%%%%
%%%%%%%%%%%%%%%%%%%%%%%%%%%%%%%%%%%%%%%%%%%%%%%%%%%%%%%%%%%%%%%%%%%%%%
%%%%%%%%%%%%%%%%%%%%%%%%%%%%%%%%%%%%%%%%%%%%%%%%%%%%%%%%%%%%%%%%%%%%%%
\section{Radial Velocities}
\label{RAV}
%%%%%%%%%%%%%%%%%%%%%%%%%%%%%%%%%%%%%%%%%%%%%%%%%%%%%%%%%%%%%%%%%%%%%%
%%%%%%%%%%%%%%%%%%%%%%%%%%%%%%%%%%%%%%%%%%%%%%%%%%%%%%%%%%%%%%%%%%%%%%
%%%%%%%%%%%%%%%%%%%%%%%%%%%%%%%%%%%%%%%%%%%%%%%%%%%%%%%%%%%%%%%%%%%%%%

We used a cross-correlation technique (Griffin 1967; Tonry \& Davis 1979) to 
measure the radial velocities of all our targets. The cross-correlation of the 
extracted and wavelength calibrated spectra is included in
the girBLDRS pipeline, in the last step, after the correction of the wavelength
calibration using the simultaneous calibration fibres. 

In the cross-correlation technique the observed spectrum is matched
(correlated) with the template. The observed and template spectra
should be of a similar spectral type in order to minimize systematic
errors due to template missmatch. The template spectrum is shifted in
wavelength (velocity) space and the difference with respect to the
observation computed until the best fit is found.

The particular implementation of the cross-correlation technique
adopted in the girBLDRS is described by Baranne et al.\ (1996), and
Dubath et al.\ (1990).  It uses the numerical mask as the template for
cross correlation, in our case based on the solar spectrum from the Solar
Flux Atlas of Kurucz et al. (1984). For most solar spectral lines the
mask provides the lower and upper limits of a wavelength window
centered on the line. For any shift in wavelength between the observed
spectrum and the numerical mask, the integral of the observed spectrum
within the mask windows is calculated. 
The whole cross-correlation
function is then constructed by evaluating the integrals for shifts
ranging from a minimum to a maximum selected velocity. More
specifically, the template is shifted to a minimum radial velocity and
the integral of the observed spectrum passing through the numerical
mask evaluated. Then the template is shifted by steps of 0.005 nm (2.8
{\rm km s$^{-1}$}) up to the highest chosen radial velocity.  The
cross-correlation function is computed at each step over the whole
velocity domain.  In this way, the cross-correlation function is an
``average spectral line'' over all the lines present in the
template. It is fitted with a Gaussian, and its peak corresponds to
the wavelength (velocity) shift of the object spectrum, and therefore
it represents a direct measure of the radial velocity. The girBLDRS
pipeline computes the radial velocity for all the stars, including the
heliocentric correction. 
The width of the cross-correlation function
depends on the instrumental and object line broadening.

The larger the number of spectral lines in the considered wavelength
interval, and the higher the resolution of the spectrum, the higher is
the precision of the velocity determination.  Therefore, in order to
achieve the best possible precision, we have used the HR9 setup of the
FLAMES+GIRAFFE spectrograph, because it covers a wavelength region
very rich in iron lines, has one of the highest resolutions offered by
the instrument, and the highest throughput for the G and K-type
stars. For the cross-correlation process we selected a solar template
mask (G2), because it corresponds to the average spectral type of our
targets, given that we observed red giant branch stars with magnitudes
ranging from the tip of the RGB ($V=11$) to one magnitude below the turn
off ($V=16.8$).

In order to check whether our radial velocities are affected by
systematic errors due to template missmatch we made the following
test.  First of all, we plotted all radial velocities as a function
of the color $B-V$ and of the $V$  magnitude. Figure 5
shows that there is no evidence of any systematic dependence of the radial
velocity either with the magnitude, or with the $B-V$ color.

We computed the radial velocities for several plates also using an F0
star template mask.  In order to compare the two results, we
calculated the radial velocity differences computed by using the two
different templates: we found an average value of $65 \pm 11$ {\rm m
s$^{-1}$}.  We repeated the same test only with the few stars that we
observed in the Horizontal Branch (see Fig.~1), and we found that the mean radial
velocity difference computed using the two different templates is
about $80 \pm 34$ {\rm m s$^{-1}$}.

\begin{figure}[!ht]
\resizebox{\hsize}{!}{
\includegraphics{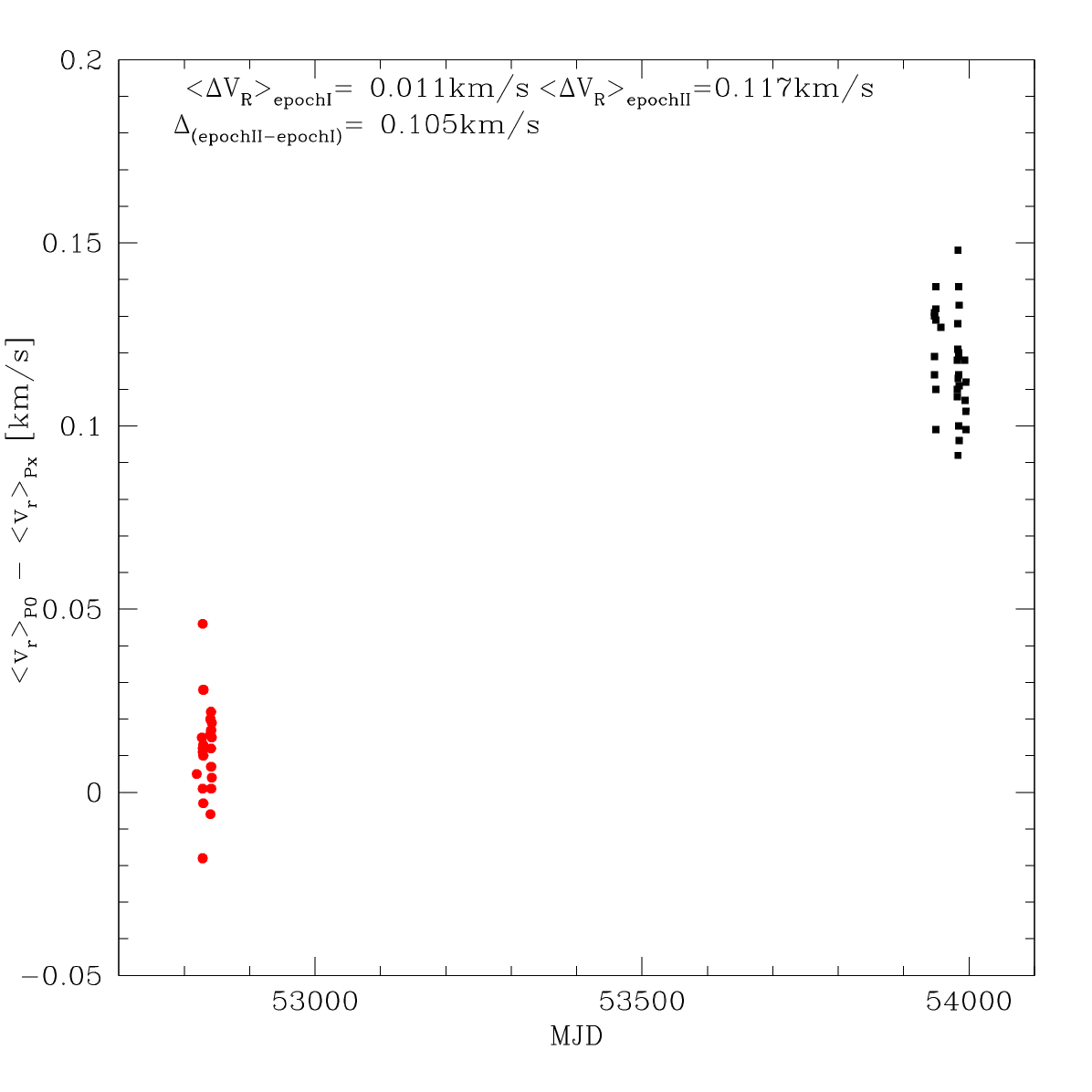}
}
\caption[]{The  average radial  velocity   differences  for  the  simultaneous
calibration  fibers between  the  reference plate 
and  all  the other  plates (Px)
plotted as a function of the Modified Julian day.  There is a systematic offset
in the radial velocity zero point between the two epochs. }
\label{DVR}
\end{figure}

%%%%%%%
\subsection{The zero point velocity shift between the epochs}
\label{ZPS}
%%%%%%%

In this section  we describe how we  minimized the plate-to-plate
and epoch-to-epoch  systematic errors  (global shifts in  the velocity
reference frame).

As described  in Sect. 3  and in Sect. 4, within each plate the
velocities were computed on the  common reference system, which --if there
were no systematic errors-- should be  the same for  all the plates,  and should
coincide with the Solar system barycentric rest-frame.

In order to measure
any possible deviations from  this, we  considered --in  each plate--  the five
simultaneous calibration fibres (described in Sect. 3).
In all our plates, these  fibres provide Th-Ar calibration-lamp spectra  
for  which we  can measure also wavelength shifts.

To test the presence and size of the average radial velocity differences between
different plates we first defined  a radial velocity system  common to all
the  plates.  This has been assumed to be the average  of the  five calibration
fiber velocities {\em of the first  plate, of the first night, and of
the first epoch}.  In the following we will refer to this plate as 
the {\em reference-plate}.
We  then   calculated  the  difference  of  the   velocities  from  the
calibration   fibres  of   other   plates,  with respect to that of   the
reference-plate.
The result is shown in Figure 6 where the amount of the global shift
in {\rm km s$^{-1}$} is plotted as a function of the Modified Julian
day.  It appears that within each epoch the velocities of plates are
registered to a common reference system within $\sim$ 50 {\rm m
s$^{-1}$} Going from epoch~I to epoch~II/epoch~III the shifts are
larger, being of the order of $\sim$100 {\rm m s$^{-1}$}.

We then corrected our  plate-to-plate, and epoch-to-epoch, global shifts
simply applying  to all plates their velocity difference with
respect to the reference plate.
This put  all the velocities from different  plates into a common
radial  velocity system  within  
$\sim 100$ {\rm m s$^{-1}$}  (not  necessarily
coincident with the true Solar system barycenter).

\begin{figure}
\resizebox{\hsize}{!}{
\includegraphics{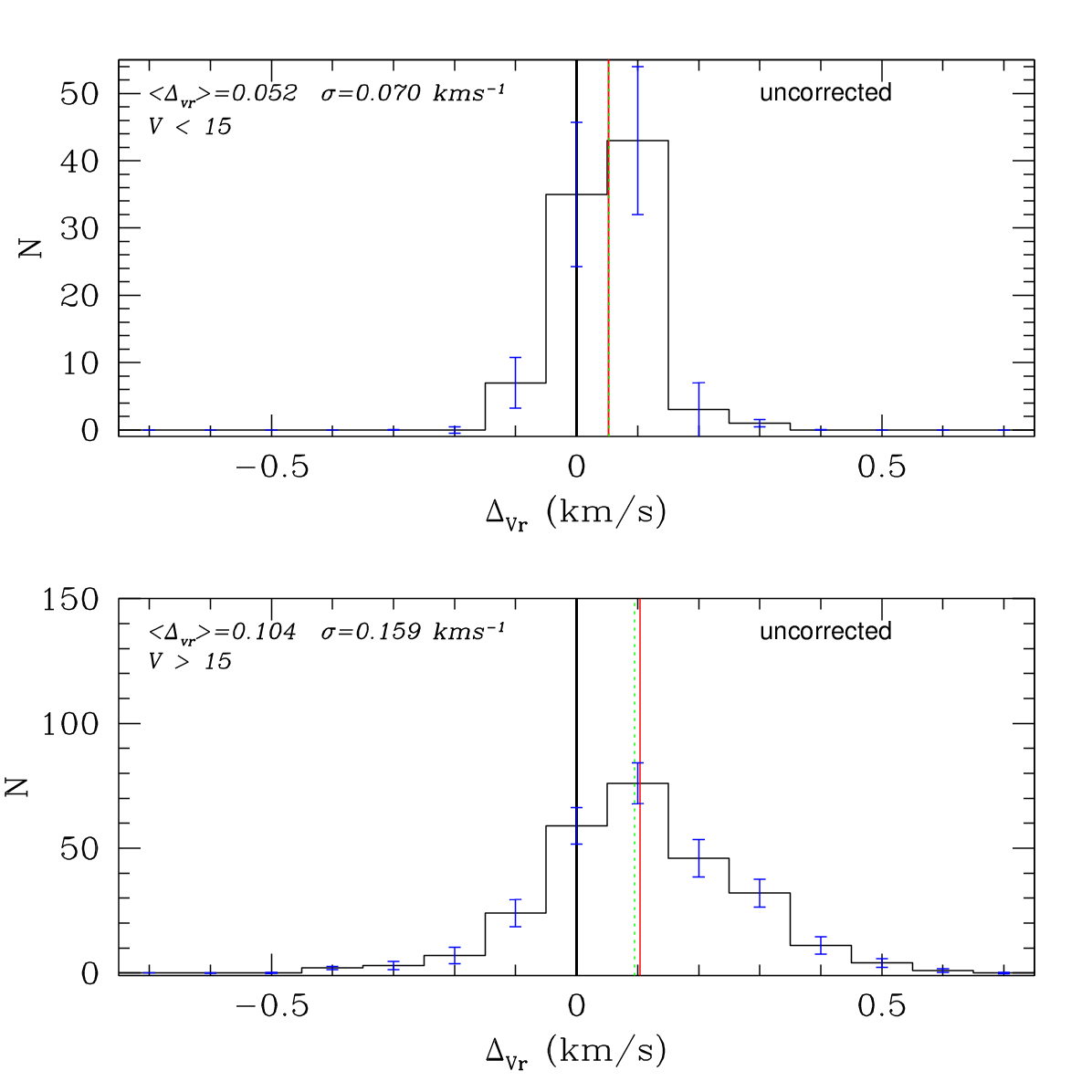}
}
\caption[]{ Histogram of differences of multiple radial velocity
measurements used to evaluate the reliability of our 
calibrations. 
In this
case no corrections have been applied to our measurements.  The upper
panel shows the 
bright stars ({\it V $<$15})
the bottom panel shows the  faint stars ({\it V $>$15}).
The solids lines are the mean, and
the dotted lines are the median.
The thick black line mark the position of a null difference. 
}
\label{WO}
\end{figure}

\begin{figure}
\resizebox{\hsize}{!}{
\includegraphics{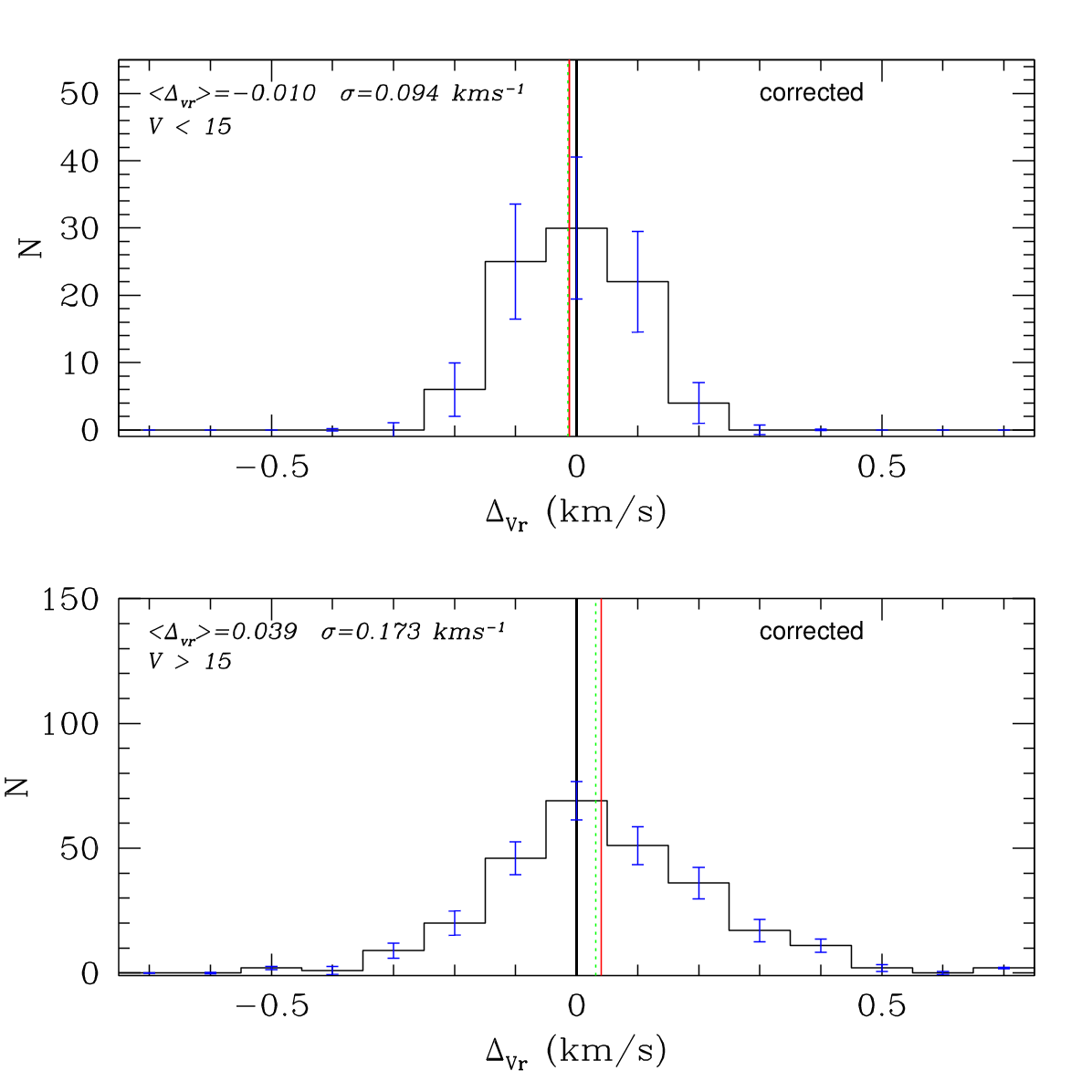}
}
\caption[]{
Histogram  of differences  of multiple radial  velocity
measurements used to evaluate the  reliability of our
calibrations. 
In this
case  the shift  found in  the  way described  in the  text has  been
applied.  In the upper panel we  show the results for the 
bright stars ({\it V}$<$15), the  bottom panel  for the 
faint  stars ({\it V}$>$15).
The solid lines are the mean, and the dotted lines are the median. 
The thick black line mark the position of a null difference. 
}
\label{WI}
\end{figure}

Figures 7 and 8 show the results of application of the above procedure.
The two figures shows the differences in radial velocity for stars
with multiple spectra before (Fig.~7) and after (Fig.~8) the
application of the correction.
In the upper panel, there are the bright stars with ({\it $V<15$}),
in the bottom panel, the faint stars with ({\it $V>15$}).  In these
figures the 
the thick black line mark the position of the zero, the 
solid lines represent the mean and the dotted lines the
median.  After the subtraction of the zero point shifts the mean and
the median are close to zero, as expected.
%
%

%%%%%%%
\subsection{M4 radial velocity}

Some systematic trends in our radial velocities are  expected.    Effects  such  as
gravitational redshift ($\sim$0.5 {\rm m s$^{-1}$} for red dwarf stars) and
convective blue shifts ($\sim-$0.2-$-0.3$ Madsen et al.\ 2002) {\rm km s$^{-1}$} for Red Giants) 
are  not affecting differential  measurements for the binary  search,
but they  do affect the  absolute value of  the velocity in  the Solar
system barycenter rest-frame.
We did  not attempt any correction  for them, since  these two effects
are  pushing   in  opposite  direction,  we  simply   add  an  average
contribution of 0.3 {\rm km s$^{-1}$} to our uncertainties on the
absolute radial velocities.
Moreover, because of the zero point global shifts mentioned in Section
4.1, we have an additional systematic error of $\sim$0.1 {\rm km s$^{-1}$}. 
In conclusion, the resulting average velocity for M4 is:
$$\overline{{\rm V_r}} = 70.29 ~ \pm 0.07 ~ (\pm 0.3) ~ (\pm 0.1) ~ {\rm ~ km ~ s^{-1}}$$
where the first  source of of uncertainties is  our estimated internal
error.   

%%%
%   FIGURE 
%__________________________________________________ 
%
\begin{figure}[!ht]
\resizebox{\hsize}{!}{ 
\includegraphics{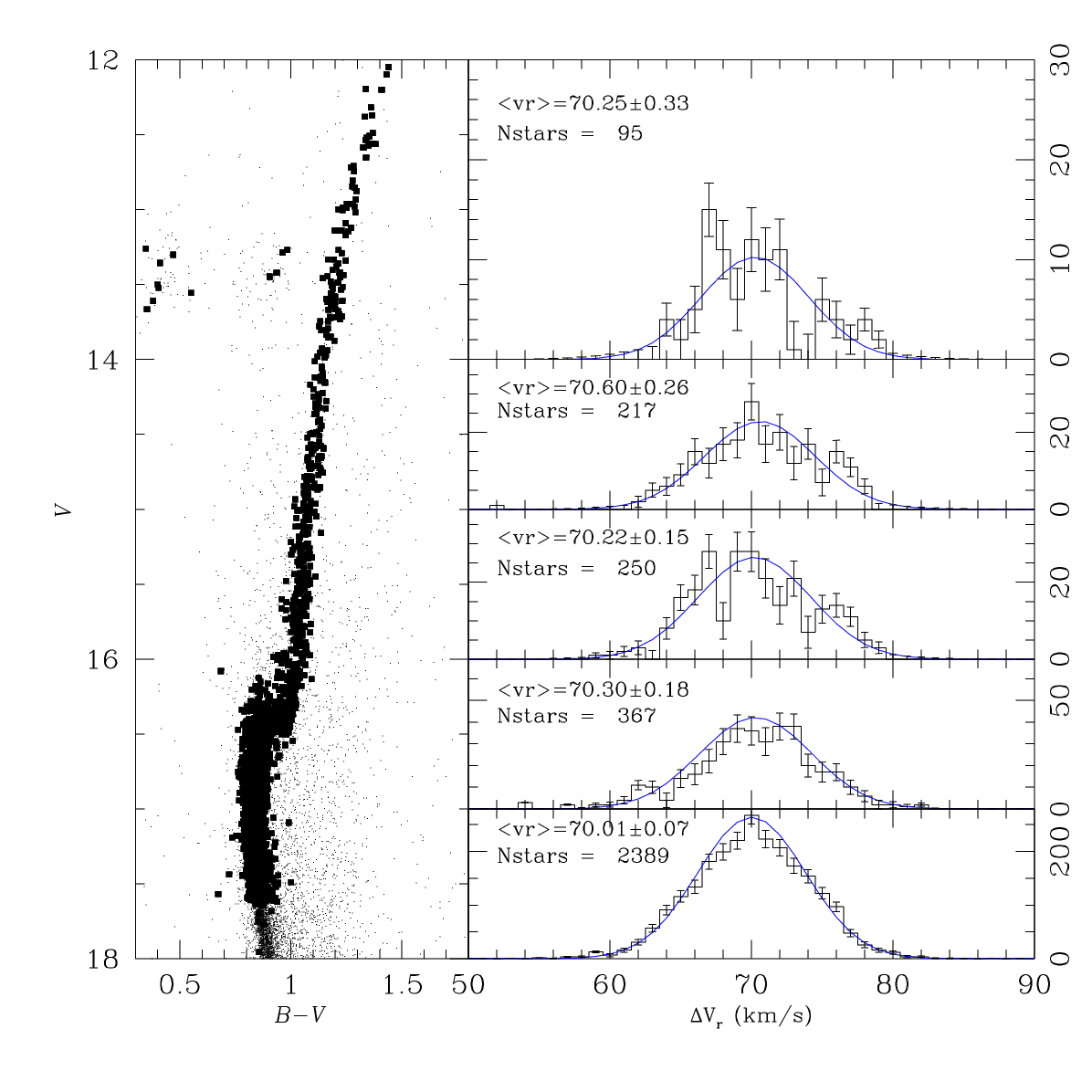}}
\caption[]{
{\em Left: } Color magnitude diagram of the stars in our photometric catalog. 
Thick dots indicate the target stars for which we have  spectra. 
{\em Right: } For five different magnitude intervals, the histogram of
the radial  velocity distribution.  
Note that, in order to increase the statistics, the brightest bin cover a 
wider range in magnitude. 
}
\label{gauss}
\end{figure}
%__________________________________________________ 
%

The left  side of Fig.~9 shows the CMD for
our target stars. We divided  our sample of  stars in
five bins  in {\it $V$} magnitude.  The right panel of
Fig.~9 shows the histogram of the distribution in radial
velocity ${\rm V_r}$ of the stars in each bin, and the least
square best fit Gaussian.

The labels in each panel give the values of the average radial
velocity ($\overline{{\rm V_r}}$) for that bin and its 
dispersion ($\sigma_{\overline{{\rm V_r}}}$).

The weighted mean of the five independent values is 
$$<~\overline{{\rm V_r}}~> ~ = ~70.27 \pm 0.19 ~ {\rm ~ km ~ s^{-1}}$$
in perfect agreement with the value obtained above. 

Our measured radial velocity is in   good agreement  with previous  studies.
In particular, Peterson et al.\  (1995) measured  the average radial velocity
of M4 to be ${\rm V_r} = 70.9\pm0.6$ {\rm km s$^{-1}$} from  the radial velocities of
200  giant stars. C\^ot\'e  \& Fisher  (1996) analyzed  33 Turn-off
dwarf stars and reported a value of $70.3\pm0.7$ {\rm km s$^{-1}$}.

\section{Radial velocity variations, and binary candidates}

In this section we describe the selection of the binary candidates
from the measured radial velocity variations.

Our approach, which  is adequately good for stars with
both two or more 
epochs, is  based on  the comparison  of  the observed
weighted rms velocity  for  single stars  with the average rms  for each star at the
same magnitude.  Stars  with rms    several times  larger   than
expected are likely to be binary candidates.

The weight $w_{i,j}$ for the $i$-star, observed in the $j$-plate used
to compute the rms, is provided by the pipeline formal errors
$\sigma_{i,j}$, according to the relation:
$w_{i,j}=1/\sigma_{i,j}^{2}$.  
We note that these errors underestimate the true errors, but they contain 
the valuable 
information on the goodness of the Gaussian fit to the
cross-correlation function, which is related to the goodness of the
observed spectrum (see Sect.3).  More explicitly, this strategy turns
out to be the most robust against false alarms caused by a bad
spectrum (which received an almost null weight), i.e. spectra with
particularly low SNR, or altered by cosmic ray events.
The weighted mean radial velocity for the $i$-star observed in $n$
plates, labeled as $j$-plate, with $j$ from 1 to $n$ (with $n=$2,
3, or 4, depending on the available number of epochs) has been
calculated as:
$$ \overline{\rm {\rm V_{r;{\it i}}}} = 
\frac{ 
   \Sigma_{j=1}^{n} w_{i,j} {\rm V_{r;{\it i,j}}}
}{
   \Sigma_{j=1}^{n} w_{i,j}
}. 
$$
Accordingly, we could define a {\em weighted radial velocity rms} as: %
$$ 
(\sigma_{ \overline{\rm {\rm V_{r;{\it i}}}}})^{2} = 
\frac{ 
 \Sigma_{j=1}^{n} w_{i,j} ~ ({\rm V_{r;{\it i,j}}} - \overline{\rm {\rm V_{r;{\it i}}}})^{2} 
}{
 \Sigma_{j=1}^{n} w_{i,j}
}. 
$$

However, for small population sample --such as in our case-- it is
customary to use an unbiased estimator for the population variance. In
normal unweighted samples, the ($n$) in the denominator (corresponding
to the  sample size) is changed  to ($n-1$).  While this  is simple in
unweighted samples, it becomes tedious for weighted samples. Thus, the
unbiased estimator of weighted population variance is given by:
$$ 
(\sigma_{ \overline{\rm {\rm V_{r;{\it i}}}}})^{2} =
\frac{
\Sigma_{j=1}^{n} w_{i,j}
}{
(\Sigma_{j=1}^{n} w_{i,j})^{2} -  \Sigma_{j=1}^{n} w_{i,j}^{2}
}
~ \Sigma_{j=1}^{n} w_{i,j} ({\rm V_{r;{\it i,j}}} -  \overline{\rm {\rm V_{r;{\it i}}}})^{2}
$$

\begin{figure}[!ht]
\resizebox{\hsize}{!}{\includegraphics{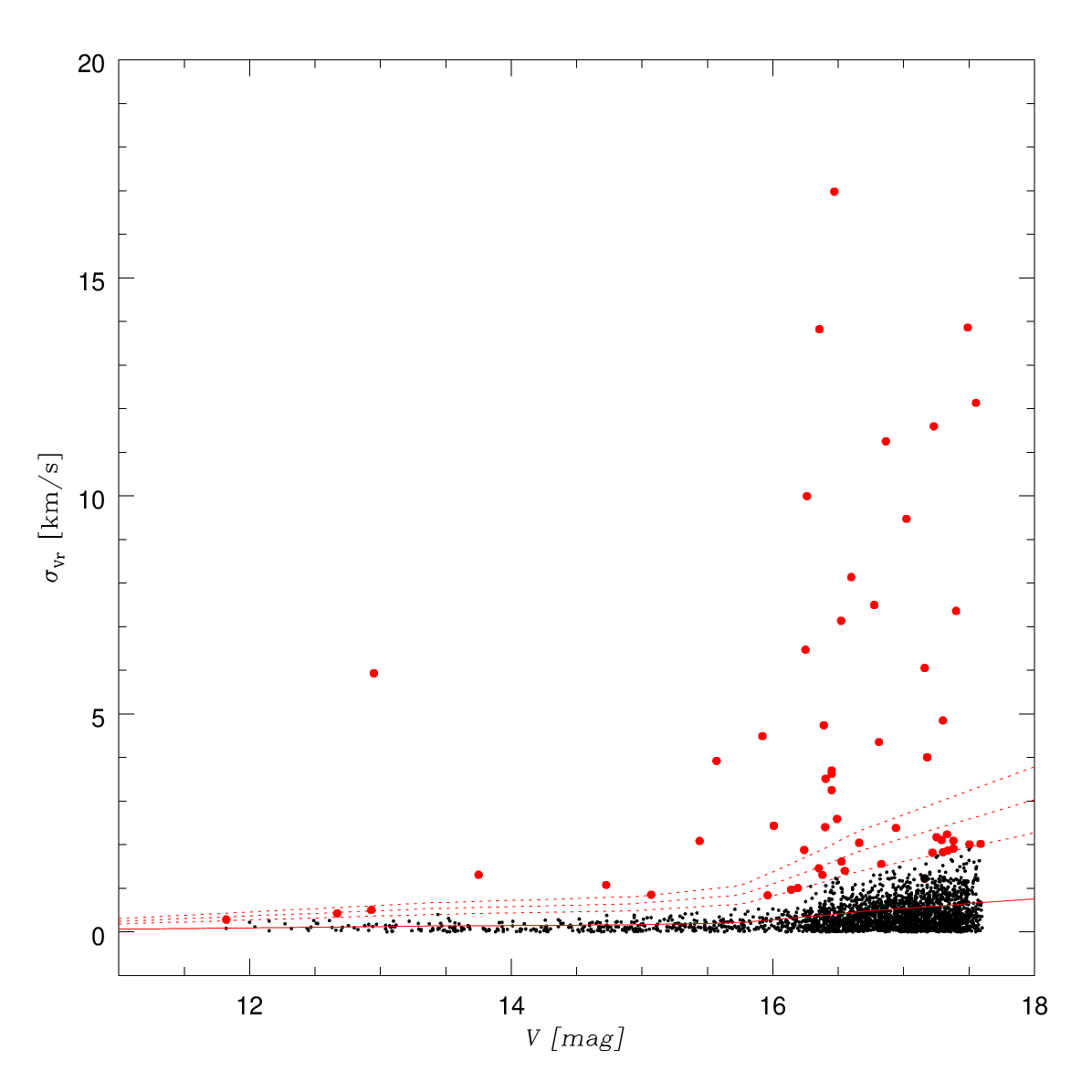}}
\caption[]{
The weighted rms as a function of the magnitude is shown for all the 
M4 stars which have been  observed in at least two epochs.
A large rms is  likely to be due to an intrinsic variation of
the radial  velocity. The solid  line 
connecting  $<\sigma_{\overline{\rm {\rm V_{r}}}}>$ values
computed in each magnitude bin 
 is an  indication  of  the precision  of  the radial  velocity
measurement  for  non-variable   stars.   The  dotted  lines  indicate the
3,  4, and 5~$<\sigma_{\overline{\rm  V_r}}>$ level.
The candidate binaries  are  indicated with
larger (red) dots.
}
\label{RMSVR}
\end{figure}

For each star with multiple observations we compute this quantity and
plot  it in Fig.~\ref{RMSVR}  as a  function of its magnitude.
The  majority of  the  objects occupy  a  well defined  locus on  this
observational  plane.  
We divided the stars in Fig.~10
in 6 equally populated magnitude bins, with $\sim 400$ stars in each bin.
For each one could compute the median and naively think that it would be 
representative of the average radial velocity error $<(\sigma_{\rm {\overline{V}_r}})>$
for that magnitude bin. 
However, although the observed distributions are satisfactorily
fitted with a Gaussian (with standard deviation $\sigma$), the median 
of the error distribution is not a good estimate of the sigma. 
The relation between the median and the $\sigma$ can be easily computed, 
and obtain the final  $<(\sigma_{\rm {\overline{V}_r}})>$ as 0.675~$\sigma$. 
Therefore, to avoid to underestimate the errors, we applied this
correction at the value of the median. 
We note, that it would be also possible to estimate
the $\sigma$ from the root mean square value of the $<\sigma_{\overline{\rm {\rm V_{r,{\it i}}}}}>$
but this estimate is too sensitive to the outliers, most of which are
binary candidates.  
The operation has been repeated twice to converge on the value of the
median, and so of the $\sigma$; in the second step, all stars
deviating by more than 3.5$<\sigma_{\overline{\rm {\rm V_{r}}}}>$
from the first median estimate have been removed.
The 1-$<\sigma_{\overline{\rm {\rm V_{r}}}}>$ value as a function of the
magnitude is indicated with a solid thin-red-line in Fig.~10. 
This is the simple connection of the  $<\sigma_{\overline{\rm {\rm V_{r}}}}>$
values computed in each magnitude bin.
The dotted lines indicate the 3, 4, and 5-$<\sigma_{\overline{\rm {\rm V_{r}}}}>$
level.
For the rest of the paper we use $\sigma$ to denote $<\sigma_{\overline{\rm {\rm V_{r}}}}>$.
All the stars above the 3-$\sigma$ level
shall be considered as binary candidates in the following discussion.
They are also indicated as larger (red) dots in Fig.~10.

%%%
%   FIGURE
%__________________________________________________
%
%__________________________________________________
%

%%%%%%%%%%%%%%%%%%%%%%
%
\section{Binary fraction}

The present data allow us to identify at 3$\sigma$ level binary candidates with
velocity variation greater than $\sim0.3$ {\rm km s$^{-1}$}for the
bright stars ($V\leq15$), and velocity variations greater than $\sim0.5$
{\rm km s$^{-1}$} for the faint stars ($V>15$). We
found 57 binary candidates.  Figure~11  shows the location of the
binary candidates in the CMD. With the triangle we shown the binary candidate found
also in the photometric binary sample of Milone et al. (2008).

In Table 4 we give the list of the binary system candidates coming
from two epochs, three epochs, and in one case with four epoch
spectra. For each star, the table gives the radial velocity as measured in
the different epochs with the corresponding errors, the coordinates (RA and
DEC) in degrees, the Modified Julian Date of the observations, the number of epochs
that each star has been observed, the radial distance from the center in 
arcmin, and
the $B$ and $V$ magnitudes from the calibrated photometric catalogue 
(Momany, private communication). 
Figure~12 shows the radial distribution of the binary candidates.
The total binary fraction that we found is $f=2.3\pm0.3\%$. 
The quoted the error is 1-$\sigma$ error. 
This is significantly smaller than the binary fraction presented 
 by Cot\'e \&  Fischer 1996, but  within the erros it is consistent, since
Cot\'e \&  Fischer quote $\pm15\%$ error.
We split our sample in two subsamples on the basis of the magnitude
of the star in order to calculate the percentage of the binary
system candidates below and above the turn off.  We found that the
fraction of binaries among stars below the turn off is $f=1.9\pm0.3\%$, while
the fraction above the TO is $f=2.7\pm0.4\%$.  
We detected a total of 4 candidates out of 97 observed targets inside
the core radius, and 53 candidates out of 2372 observed stars outside
the core radius. These numbers imply a lower limit for the binary fraction
$f=4.1\pm2.1\%$ inside the cluster core, and $f=2.2\pm0.5\%$ outside the
core.
We also found that that the fraction of binaries inside the  half mass
radius ($r_{h}=2.3$pc) is $f=3.0\pm0.4\%$, while outside $1 r_h$ we
have $f=1.5\pm0.3\%$. 
It is worth noticing that the sample that we found
represents only spectroscopic binary candidates  and not the total population of
binary systems: based on Gaussian statistic it is probable that our 
sample includes some false alarms,
 and surely it is incomplete.  With at least one other epoch we could
confirm the candidates found with this survey, and reach a significantly 
higher completeness in our estimate of the binary fraction.
In order to quantify what we would gain from one additional epoch, and
to have an empirical estimate of the completeness of our present
search, we analyzed the binary 
fraction we obtained from the subsample of stars with three epochs.
We  calculated how many binary candidates have been detected using 
three epochs of the data and we compared this number with the 
candidates that would be detected using only two epochs.
We found that using only two epochs we can identify 8 candidates out
of the available 484 targets. We analyzed  all the possible 
combinations between the epochs, and the results were the same, 
and therefore robust. 
Using all three epochs we found 12 candidates out of  the 484 targets.
All the 8 candidates from the two epochs observations could be
confirmed adding the third epoch.
A third  epoch of observations is extremely important: our simple test
shows that the binary fraction from two epochs suffers from an
incompleteness of at least of $40\%$.

Accounting for this incompleteness, and considering the number of targets
with two and three epochs observed inside and outside the core,
we have that the fraction of binary candidates in the whole sample is
$f=3.0\pm0.3\%$, inside the core becomes 
$f=5.1\pm2.3\%$, and $f=3.0\pm0.4\%$ outside the core. In the same
way, we find $f=4.5\pm0.4\%$ and $f=1.8\pm0.6\%$ for the fraction of
binaries inside and outside the cluster half mass radius.

\begin{figure}[!ht]
\resizebox{\hsize}{!}{\includegraphics{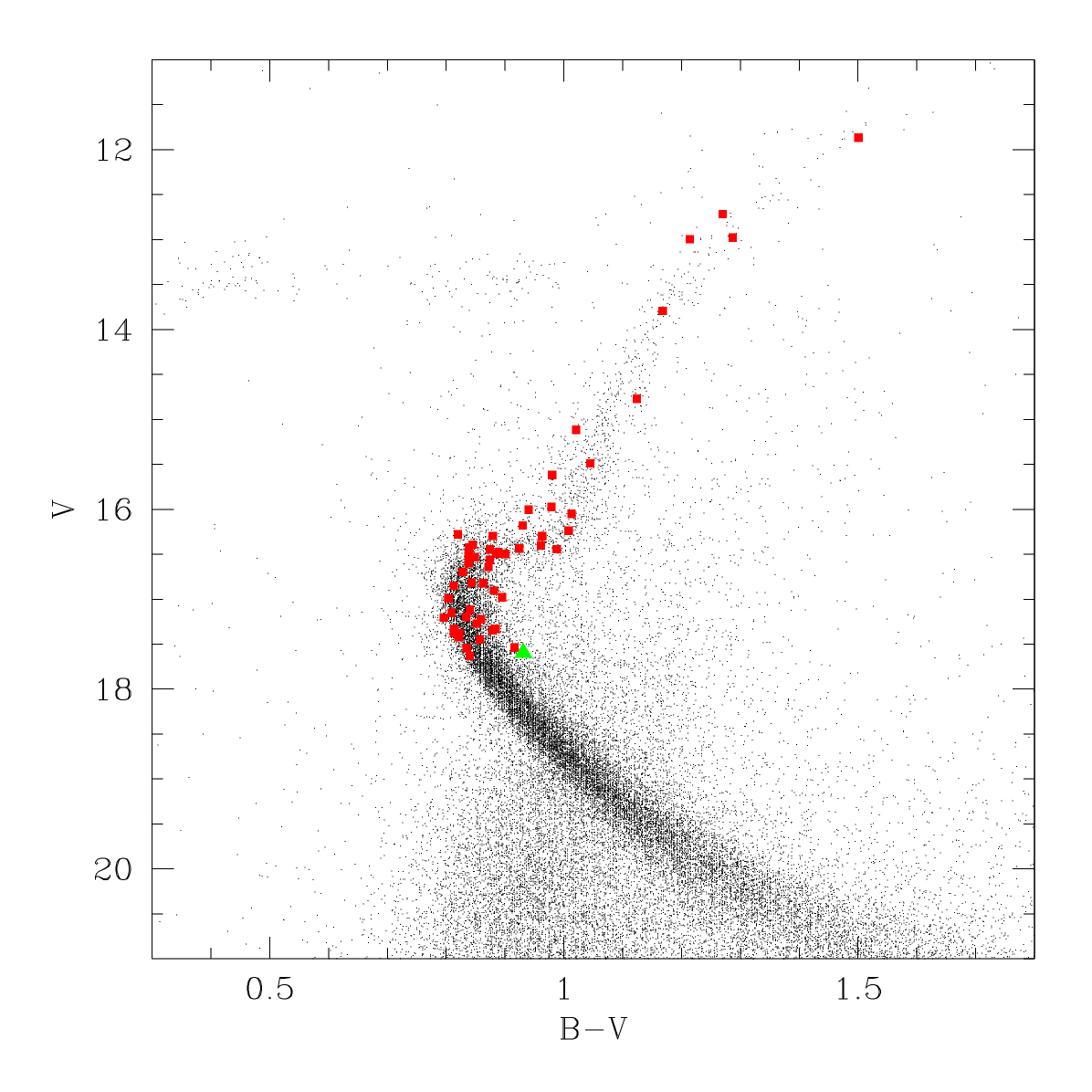}}
\caption[]{Location of the binary candidates (large red dots) in the CMD. 
The triangle (green in electronic version) indicates binary candidate found
also in the photometric sample (Milone et al. 2008).
}

\end{figure}

\begin{figure}[!ht]
\resizebox{\hsize}{!}{\includegraphics{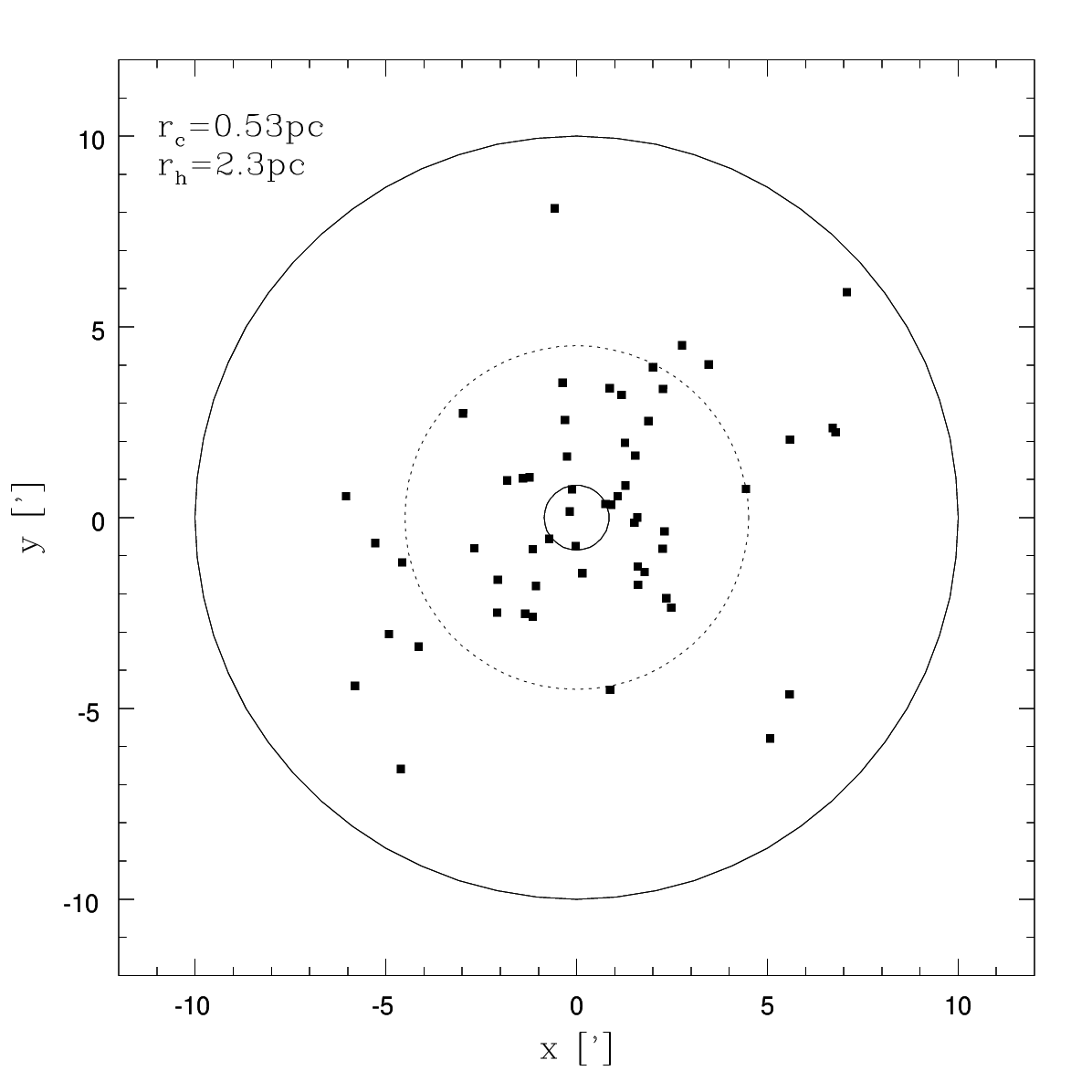}}
\caption[]{Radial distribution of the binary system candidates. 
The inner circle represents the cluster core. The outer circle 
indicates the total area of the cluster covered by observations.
The dotted circle contains the binaries inside the half mass radius.
}

\end{figure}
%%%%%%%%%%%%%%%%%%%%%%

\section{Conclusions}
We have obtained 5973 individual spectra of 2469 
stars in the Galactic globular cluster M4. Each star has
been observed at least twice, in a temporal interval of about
three years. For 484 stars we have three epochs. All the data were
obtained with the same instrument FLAMES+GIRAFFE at VLT, and with 
the same set-up. This database represents the largest multi-epoch 
high resolution spectroscopic sample ever collected in a globular 
cluster.

The observed stars cover a large range in luminosity, 
from the red giant branch tip,
to more than 1 magnitude below the TO, and cover most of the cluster extension,
from the inner core to the outskirts. 

We cross-correlated all the stellar
spectra with the solar template to derive accurate radial velocities. The average
radial velocity of the observed stars is $70.29 \pm 0.07 (\pm 0.3)(\pm0.1)${\rm km s$^{-1}$}, where the
systematic errors in the parenthesis include the effects such as gravitational
redshift, convective blue shifts, and the global zero point uncertainty of the radial
velocity. No systematic shift in average velocity is observed as a function of
the stellar magnitude and color. 

The search for variations in radial velocities among the stars with multi-epoch
observations yielded 57 binary star candidates. The candidates in the bright sample 
($V\leq 15$) have $3\sigma$ velocity 
variations larger than $\sim 0.3$~km~s$^{-1}$, while the fainter candidates
($V>15$~mag) have $3\sigma$ velocity 
variations larger than $\sim 0.5$~km~s$^{-1}$.

There are four binary star candidates out of 97 observed targets inside the core
radius, and 53 candidates out of 2372 observed stars outside the core radius. 
We have shown that increasing the stellar coverage from two epochs
to three epochs, the number of binary candidates increases by 40\%.
Accounting for this incompleteness affecting stars with only two epochs, we
have found that the lower limit for the binary fraction is
$f=3.0\pm0.3\%$, the binary fraction in the cluster core is 
$f=5.1\pm2.3\%$, which decreases to $f=3.0\pm0.4\%$ outside the core. 
Similarly, we found $f=4.5\pm0.4\%$ and $f=1.8\pm0.6\%$ for the binary fraction 
inside and outside the half mass radius. As
expected from energy equipartition, the binary fraction is higher
in the central parts of the cluster.

In the next paper we will present detailed completeness simulations to constrain 
accurately the total binary fraction in the cluster as a function of magnitude, period,
and other orbital parameters. These will then be used together with the dynamical
evolutionary modeling of Heggie \& Giersz (2008) to derive the limits for the
primordial binary fraction in M4.\\
This model is specifically tailored to this cluster.  It includes all
the main processes affecting the distribution of single stars and
binaries in 12Gyr of evolution, i.e. two-, three- and four-body
encounters, the galactic tide, and the internal evolution of single
and binary stars.  It provides satisfactory fits to the
surface-brightness profile, velocity dispersion profile and luminosity
functions (in two fields) of M4, but the properties of its binary
stars are still poorly constrained.  We shall examine this model using
the same observational constraints (on magnitude, radial distribution
and cadence) as in the actual observations of M4 reported in this
paper, in order to test our estimates of the completeness of the
observational sample, and to improve our assumptions on the initial
abundance and other parameters of the binaries in this cluster.

%%%%%%%%%%%%%%%%%%%%%%%%%%%%%%%%%%%%%%%%%%%%%%%%%%%%%%%%%%%%%%%%%%%%%%
%%%%%%%%%%%%%%%%%%%%%%%%%%%%%%%%%%%%%%%%%%%%%%%%%%%%%%%%%%%%%%%%%%%%%%
%%%%%%%%%%%%%%%%%%%%%%%%%%%%%%%%%%%%%%%%%%%%%%%%%%%%%%%%%%%%%%%%%%%%%%

\begin{acknowledgements}
We thank Claudio Melo for his kind advice on the 
pipeline and Antonino Milone for providing his
reults on the photometric binary. We also thank 
the referee for his helpful comments and suggestions. 
\end{acknowledgements}

----------------------------------------
\begin{table*}
\caption{List of the binary system candidates found. The columns are: ID, radial velocity and 
relative error that comes form the pipeline, the mean of the radial velocity, the weighed error,  
the  Modified Julian Day of observations, number of epochs,
coordinates (RA and DEC), distance from the center in arcminutes, \it{V}, \it{B} magnitude. }
\begin{center}

\begin{tabular}{lcccccccccccc} \hline
\hline
ID    & ${\rm {\rm V_{r;{\it i,j}}}}$                   
               &  $\sigma_{\rm {\rm V_{r;{\it i}}}}$-pipe. 
                       & $\overline{\rm {\rm V_{r;{\it i}}}}$
                               & $\sigma_{\overline{\rm {\rm V_{r;{\it i}}}}}$
                                       & MJD
                                                    & $n$ 
                                                        &  R.A.$_{J2000.0}[^\circ]$      
                                                                     & Decl.$_{J2000.0}[^\circ]$      
                                                                                 &  d[$'$]  &  $V$    &  $B$    \\
 & [km/s] & [km/s] & [km/s] & [km/s] & & & & & & & \\
%%%%
\hline
\hline\\
20979 &  72.37 &  0.41 & 70.09 & 2.23  & 53933.93 & 2 &  245.8123  &  -26.5450 &  4.742 &  17.378 &  18.191 \\ 
      &  67.82 &  0.34 &       &       & 52827.18 &   &      	     &           &        &         &         \\
26688 &  85.41 &  0.28 & 66.43 & 11.25 & 52828.14 & 3 &  245.9139  &  -26.6006 &  4.605 &  16.865 &  17.788 \\
      &  63.03 &  0.28 &       &       & 53983.10 &   &	     &           &        &         &         \\
      &  50.87 &  0.28 &       &       & 53947.09 &   &	     &           &        &         &         \\
28573 &  76.79 &  0.30 & 66.22 & 13.86 & 52841.20 & 2 &  245.8724  &  -26.5675 &  2.881 &  17.490 &  18.45  \\
      &	 47.65 &  0.42 &       &       & 53994.19 &   &	     &           &        &         &         \\
31674 &  61.22 &  0.18 & 63.71 & 2.59  & 52819.11 & 2 &  245.9404  &  -26.5315 &  2.314 &  16.532 &  17.380 \\ 
      &  66.20 &  0.01 &       &       & 53982.97 &   &	     &           &        &         &         \\
29065 &  66.20 &  0.01 & 66.72 & 0.50  & 52819.11 & 2 &  245.9413  &  -26.5607 &  3.157 &  12.979 &  14.266 \\
      &  67.24 &  0.01 &       &       & 53984.09 &   &	     &           &        &         &         \\
29725 &  73.13 &  0.22 & 81.81 & 6.47  & 52819.11 & 2 &  245.8590  &  -26.5526 &  2.655 &  16.296 &  17.175 \\ 
      &  90.50 &  0.50 &       &       & 53983.03 &   &	     &           &        &         &         \\
33538 &  72.78 &  0.20 & 74.12 & 1.30  & 52829.23 & 2 &  245.9802  &  -26.5130 &  4.478 &  16.426 &  17.265 \\ 
      &  75.46 &  0.25 &       &       & 53983.03 &   &	     &           &        &         &         \\
34848 &  66.75 &  0.12 & 65.53 & 1.07  & 52822.21 & 2 &  245.8927  &  -26.4988 &  1.608 &  14.771 &  15.895 \\ 
      &  64.31 &  0.07 &       &       & 52982.99 &   &	     &           &        &         &         \\ 
36750 &  69.10 &  0.12 & 71.56 & 2.43  & 52822.21 & 2 &  245.9194  &  -26.4718 &  3.405 &  16.051 &  17.064 \\ 
      &  74.03 &  0.14 &       &       & 53983.08 &   &	     &           &        &         &         \\ 
32874 &  67.79 &  0.02 & 68.24 & 0.42  & 52822.21 & 2 &  245.9145  &  -26.5198 &  0.948 &  12.717 &  13.987 \\ 
      &  68.70 &  0.01 &       &       & 53984.09 &   &	     &           &        &         &         \\
31015 &  67.17 &  0.02 & 75.29 & 5.93  & 52822.21  & 3 &  245.8970  &  -26.5379 &  0.761 &  12.997 &  14.211 \\ 
      &  79.30 &  0.01 &       &       & 53983.05 &   &	     &           &        &         &         \\
      &  79.40 &  0.01 &       &       & 53983.08 &   &	     &           &        &         &         \\
35327 &  71.21 &  0.28 & 72.43 & 1.22  & 52827.15 & 2 &  245.9210  &  -26.4928 &  2.308 &  17.201 &  18.035 \\ 
      &  73.65 &  0.29 &       &       & 53984.98 &   &	     &           &        &         &         \\
5195  &  62.88 &  0.36 & 64.10 & 1.32  & 52840.14 & 2 &  245.8685  &  -26.3794 &  8.892 &  17.261 &  18.118 \\ 
      &  65.32 &  0.31 &       &       & 53995.05 &   &      	     &           &        &         &         \\
37894 &  76.21 &  0.32 & 78.12 & 1.91  & 52840.14 & 2 &  245.9490  &  -26.4502 &  5.271 &  17.421 &  18.242 \\ 
      &  80.03 &  0.32 &       &       & 53984.06 &   &      	     &           &        &         &         \\
5001  &  67.53 &  0.44 & 69.20 & 1.55  & 52828.25 & 2 &  245.8867  &  -26.3904 &  8.112 &  16.821 &  17.684 \\
      &  70.87 &  0.30 &       &       & 53982.01 &   &      	     &           &        &         &         \\
19378 &  75.17 &  0.47 & 72.44 & 2.38  & 52828.25 & 2 &  245.8117  &  -26.6353 &  8.065 &  16.992 &  17.796 \\
      &  69.71 &  0.27 &       &       & 53947.09 &   &      	     &           &                  &         \\
28771 &  73.42 & 0.22  & 65.30 & 8.13  & 52828.25 & 2 &  245.9437  &  -26.5648 &  3.418 &  16.600 &  17.513 \\
      &  57.18 & 0.20  &       &       & 53948.97 &   &      	     &           &                  &         \\
30888$^*$ &  91.20 & 0.33  & 78.49 & 12.13 & 52840.07 & 2 &  245.8761  &  -26.5393 &  1.441 &  17.553 &  18.528 \\
      &  65.78 & 0.46  &       &       & 53994.19 &   &      	     &           &                  &         \\
30917 &  59.07 & 0.18  & 66.47 & 7.13  & 52840.20 & 2 &  245.8475  &  -26.5388 &  2.824 &  16.521 &  17.437 \\
      &  73.87 & 0.23  &       &       & 53984.03 &   &      	     &           &                  &         \\
36036 &  82.19 & 0.06  & 80.07 & 2.08  & 52791.08 & 2 &  245.8917  &  -26.4829 &  2.562 &  15.488 &  16.533 \\ 
      &  77.96 & 0.07  &       &       & 53983.08 &   &      	     &           &                  &         \\
37488 &  67.01 & 0.17  & 68.43 & 1.45  & 52791.08 & 2 &  245.9620  &  -26.4586 &  5.276 &  16.393 &  17.238 \\ 
      &  69.93 & 0.17  &       &       & 53984.09 &   &            &           &                  &         \\
20288 &  61.82 & 0.22  & 65.52 & 3.70  & 53982.99 & 2 &   245.8205 &  -26.5819 &  5.372 &  16.499 &  17.399 \\ 
      &  69.23 & 0.23  &       &       & 52791.08 &   &      	     &           &                  &         \\
33901 &  58.72 & 0.17  & 62.05 & 3.24  & 52791.08 & 2 &  245.8636  &  -26.5093 &  2.072 &  16.493 &  17.378 \\ 
      &  65.38 & 0.21  &       &       & 53983.08 &   &      	     &           &        &         &         \\
48830 &  77.58 & 0.47  & 75.41 & 2.17  & 53984.06 & 2 &  246.0294  &  -26.4270 &  9.202 &  17.205 &  18.001 \\  
      &  73.24 & 0.45  &       &       & 52840.17 &   &      	     &           &        &         &         \\
46654 &  69.32 & 0.32  & 71.14 & 1.81  & 52840.07 & 2 &  246.0240  &  -26.4882 &  7.125 &  17.151 &  17.960 \\ 
      &  72.97 & 0.37  &       &       & 53947.15 &   &      	     &           &        &         &         \\
36225 &  76.42 & 0.29  & 78.55 & 2.01  & 52840.07 & 2 &  245.8420  &  -26.4800 &  4.045 &  17.630 &  18.470 \\ 
      &  80.69 & 0.41  &       &       & 53993.99 &   &            &           &        &         &         \\
46544 &  61.34 & 0.31  & 68.04 & 6.05  & 52840.07 & 2 &  246.0017  &  -26.4915 &  5.931 &  17.117 &  17.957 \\  
      &  74.74 & 0.20  &       &       & 53947.15 &   &            &           &        &         &         \\
36891 &  62.00 & 0.33  & 59.98 & 2.00  & 52840.07 & 2 &  245.9397  &  -26.4693 &  4.036 &  17.543 &  18.378 \\ 
      &  57.97 & 0.36  &       &       & 53933.19 &   &            &           &        &         &         \\
32308 &  38.72 & 0.24  & 56.18 & 16.98 & 52791.08 & 2 &  245.9271  &  -26.5254 &  1.568 &  16.470 &  17.347 \\ 
      &  73.65 & 0.19  &       &       & 53983.03 &   &            &           &        &         &         \\
33234 &  69.51 & 0.11  & 68.64 & 0.83  & 52840.20 & 2 &  245.9176  &  -26.5161 &  1.190 &  15.960 &  16.944 \\ 
      &  67.77 & 0.16  &       &       & 53983.08 &   &            &           &        &         &         \\
35997 &  74.88 &  0.21 & 76.92 & 1.86  & 52841.07 & 2 &  245.9326  &  -26.4834 &  3.127 &  17.381 &  18.203 \\ 
      &  78.96 &  0.33 &       &       & 53984.06 &   &            &           &        &         &         \\

  \hline
\hline
\multicolumn{12}{l}{$^*$ Binary candidate in common with the photometric binary sample of Milone et al. (2008)}\\
\end{tabular}
\end{center}   
\end{table*}

\addtocounter{table}{-1}
\begin{table*}
\caption{List of the binary system candidates found. The columns are: ID, radial velocity and relative error that comes form the pipeline, 
the mean of the radial velocity, the weighed error,  Modified Julian Day of observations, number of epoch,
coordinates (RA and DEC), distance from the center in arcminutes, \it{V}, \it{B} magnitude.}

\begin{center}

\begin{tabular}{lcccccccccccc}
\hline
\hline
ID    & ${\rm {\rm V_{r;{\it i,j}}}}$                   
               &  $\sigma_{\rm {\rm V_{r;{\it i}}}}$-pipe. 
                       & $\overline{\rm {\rm V_{r;{\it i}}}}$
                               & $\sigma_{\overline{\rm {\rm V_{r;{\it i}}}}}$
                                       & MJD
                                                    & $n$ 
                                                        &  R.A.$_{J2000.0}[^\circ]$      
                                                                     & Decl.$_{J2000.0}[^\circ]$      
                                                                                 &  d[$'$]  &  $V$    &  $B$    \\
 & [km/s] & [km/s] & [km/s] & [km/s] & & & & & & & \\
\hline
\hline

33700 &  64.49 &  0.49 & 61.14 & 7.49  & 52828.23 & 3 &  245.9214  &  -26.5115 &  1.507 &  16.775 &  17.661 \\
      &  68.48 &  0.19 &       &       & 53983.10 &   &            &           &        &         &         \\
      &  50.47 &  0.36 &       &       & 53947.09 &   &            &           &        &         &         \\
37047 &  69.26 &  0.16 & 83.26 & 13.82 & 52822.21 & 2 &  245.8907  &  -26.4666 &  3.539 &  16.360 &  17.361 \\
      &  97.26 &  0.14 &       &       & 53983.03 &   &            &           &        &         &         \\
30205 &  71.34 &  0.22 & 66.55 & 4.00  & 52841.07 & 2 &  245.9273  &  -26.5468 &  2.043 &  17.224 &  18.083 \\ 
      &  61.76 &  0.41 &       &       & 53985.04 &   &            &           &        &         &         \\
29506 &  74.46 &  0.26 & 72.49 & 1.82  & 52841.07 & 2 &  245.8777  &  -26.5553 &  2.104 &  17.345 &  18.224 \\ 
      &  70.52 &  0.39 &       &       & 53933.99 &   &            &           &        &         &         \\
32568 &  74.72 &  0.07 & 73.86 & 0.85  & 52829.20 & 2 &  245.8941  &  -26.5228 &  0.249 &  15.115 &  16.136 \\ 
      &  73.01 &  0.07 &       &       & 53982.97 &   &            &           &        &         &         \\
31322 &  60.57 &  0.22 & 62.44 & 1.87  & 52829.20 & 2 &  245.8841  &  -26.5348 &  0.935 &  16.279 &  17.099 \\ 
      &  64.32 &  0.22 &       &       & 53983.03 &   &            &           &        &         &         \\
46733 &  69.50 &  0.35 & 67.40 & 2.08  & 53984.06 & 2 &  246.0225  &  -26.4863 &  7.085 &  17.331 &  18.144 \\ 
      &  65.31 &  0.38 &       &       & 52842.08 &   &            &           &        &         &         \\
30901 &  74.79 &  0.20 & 61.91 & 11.60 & 52842.08 & 2 &  245.9395  &  -26.5391 &  2.383 &  17.271 &  18.123 \\ 
      &  49.03 &  0.31 &       &       & 53994.99 &   &            &           &        &         &         \\
42302 &  70.38 &  0.20 & 72.02 & 1.61  & 52840.20 & 2 &  246.0015  &  -26.6027 &  7.249 &  16.477 &  17.366 \\ 
      &  73.66 &  0.25 &       &       & 53984.09 &   &            &           &        &         &         \\
29545 &  70.52 &  0.02 & 69.21 & 1.30  & 53984.03 & 2 &  245.9274  &  -26.5548 &  2.378 &  13.793 &  14.961 \\ 
      &  67.90 &  0.02 &       &       & 52829.18 &   &            &           &        &         &         \\
29951 &  61.99 &  0.15 & 62.97 & 0.96  & 52829.18 & 2 &  245.9003  &  -26.5497 &  1.475 &  16.180 &  17.110 \\ 
      &  63.96 &  0.18 &       &       & 53983.03 &   &            &           &        &         &         \\
34034 &  52.95 &  0.17 & 62.95 & 9.99  & 52829.18 & 2 &  245.8745  &  -26.5079 &  1.630 &  16.299 &  17.262 \\ 
      &  72.95 &  0.16 &       &       & 53983.08 &   &            &           &        &         &         \\
37428 &  67.71 &  0.19 & 72.07 & 4.35  & 52828.20 & 2 &  245.9348  &  -26.4598 &  4.400 &  16.849 &  17.662 \\ 
      &  76.43 &  0.18 &       &       & 53949.07 &   &            &           &        &         &         \\
34880 &  72.23 &  0.26 & 74.63 & 2.10  & 52841.07 & 2 &  245.9261  &  -26.4984 &  2.213 &  17.327 &  18.211 \\ 
      &  77.03 &  0.44 &       &       & 53994.19 &   &            &           &        &         &         \\
28609 &  81.53 &  0.27 & 74.16 & 7.36  & 52841.11 & 2 &  245.8587  &  -26.5670 &  3.274 &  17.447 &  18.304 \\ 
      &  66.80 &  0.28 &       &       & 53994.19 &   &            &           &        &         &         \\ 
19968 &  63.65 &  0.50 & 69.44 & 4.84  & 53933.99 & 2 &  245.7893  &  -26.5990 &  7.316 &  17.349 &  18.163 \\ 
      &  75.23 &  0.93 &       &       & 52841.11 &   &            &           &        &         &         \\
41610 &  70.52 &  0.43 & 81.73 & 9.47  & 52842.15 & 2 &  245.9920  &  -26.6219 &  7.692 &  17.020 &  17.874 \\
      &  92.95 &  0.35 &       &       & 53948.97 &   &            &           &        &         &         \\
29998 &  72.02 &  0.14 & 77.28 & 3.62  & 52819.11 & 3 &  245.9307  &  -26.5492 &  2.273 &  16.495 &  17.395 \\  
      &  79.26 &  0.14 &       &       & 53947.06 &   &            &           &        &         &         \\
      &  80.57 &  0.18 &       &       & 53983.98 &   &            &           &        &         &         \\
21175 &  71.67 &  0.18 & 65.69 & 4.73  & 52819.11 & 3 &  245.7992  &  -26.5366 &  5.340 &  16.437 &  17.361 \\
      &  62.03 &  0.15 &       &       & 53981.98 &   &            &           &        &         &         \\
      &  63.37 &  0.16 &       &       & 53982.97 &   &            &           &        &         &          \\
28498 &  84.03 &  0.15 & 81.89 & 2.40  & 53984.09 & 3 &  245.8761  &  -26.5688 &  2.863 &  16.443 &  17.431  \\
      &  78.42 &  0.15 &       &       & 52829.23 &   &            &           &        &         &          \\
      &  83.22 &  0.14 &       &       & 53981.98 &   &            &           &        &         &          \\
36912 &  69.93 &  0.11 & 65.47 & 3.51  & 52822.21 & 3 &  245.9136  &  -26.4690 &  3.480 &  16.447 &  17.321  \\
      &  62.64 &  0.15 &       &       & 53981.98 &   &            &           &        &         &          \\
      &  63.86 &  0.22 &       &       & 53984.09 &   &            &           &        &         &          \\
20390 &  61.05 &  0.23 & 62.98 & 1.40  & 52842.18 & 3 &  245.8059  &  -26.5764 &  5.812 &  16.598 &  17.436  \\
      &  63.93 &  0.24 &       &       & 53983.10 &   &            &           &        &         &          \\
      &  63.93 &  0.18 &       &       & 53947.09 &   &            &           &        &         &          \\
32898 &  63.18 &  0.23 & 62.62 & 2.04  & 53983.10 & 3 &  245.9117  &  -26.5195 &  0.816 &  16.698 &  17.526 \\
      &  60.26 &  0.23 &       &       & 52828.23 &   &            &           &        &         &         \\
      &  64.42 &  0.19 &       &       & 53947.09 &   &            &           &        &         &         \\
34006 &  64.02 &  0.00 & 64.07 & 0.28  &  53983.05 & 3 &  245.8713  &  -26.5082 &  1.755 &  11.865 &  13.366 \\
      &  63.82 &  0.00 &       &       & 52829.20 &   &            &           &        &         &         \\
      &  64.39 &  0.00 &       &       & 53984.09 &   &            &           &        &         &         \\
32057 &  78.73 &  0.08 & 74.44 & 3.92  & 52829.26 & 3 &  245.9256  &  -26.5277 &  1.496 &  15.618 &  16.598 \\
      &  70.78 &  0.09 &       &       & 53983.99 &   &            &           &        &         &         \\
      &  70.82 &  0.07 &       &       & 53981.98 &   &            &           &        &         &         \\
21652 &  68.49 &  0.13 & 74.80 & 4.48  & 52829.26 & 4 &  245.7849  &  -26.5161 &  6.089 &  15.973 &  16.952 \\  
      &  77.49 &  0.14 &       &       & 53947.06 &   &            &           &        &         &         \\
      &  77.71 &  0.16 &       &       & 53957.01 &   &            &           &        &         &         \\
      &  75.07 &  0.17 &       &       & 53983.98 &   &            &           &        &         &         \\
33529 & 66.55  &  0.10 & 70.40 & 1.00  & 52840.20 & 3 &  245.8953  &  -26.5131 &   0.739 &  16.238 &  17.24 \\      
      & 64.52  &  0.08 &       &       & 53947.03 &   &            &             &              &          \\ 
      & 80.14  &  0.09 &       &       & 53983.98 &   &            &             &              &          \\
\hline
\hline
\end{tabular}
\end{center}
\end{table*}

%%%%%%%%%%%%%%%%%%%%%%%%%%%%%%%%%%%%%%%%%%%%%%%%%%%%%%%%%%%%%%%%%%%%%%
%%%%%%%%%%%%%%%%%%%%%%%%%%%%%%%%%%%%%%%%%%%%%%%%%%%%%%%%%%%%%%%%%%%%%%
%%%%%%%%%%%%%%%%%%%%%%%%%%%%%%%%%%%%%%%%%%%%%%%%%%%%%%%%%%%%%%%%%%%%%%
%%%%%%%%%%%%%%%%%%%%%%%%%%%%%%%%%%%%%%%%%%%%%%%%%%%%%%%%%%%%%%%%%%%%%%
%%%%%%%%%%%%%%%%%%%%%%%%%%%%%%%%%%%%%%%%%%%%%%%%%%%%%%%%%%%%%%%%%%%%%%
%%%%%%%%%%%%%%%%%%%%%%%%%%%%%%%%%%%%%%%%%%%%%%%%%%%%%%%%%%%%%%%%%%%%%%
%%%%%%%%%%%%%%%%%%%%%%%%%%%%%%%%%%%%%%%%%%%%%%%%%%%%%%%%%%%%%%%%%%%%%%
%%%%%%%%%%%%%%%%%%%%%%%%%%%%%%%%%%%%%%%%%%%%%%%%%%%%%%%%%%%%%%%%%%%%%%

%
% %

%
\end{document}